\documentclass[acmsmall]{acmart}
\usepackage[most]{tcolorbox} 

\usepackage{longtable}
\usepackage{longtable, booktabs, tabularx, multirow} 
\usepackage{soul}

\usepackage{amssymb}

\usepackage{tikz}
\usetikzlibrary{arrows.meta, shapes, calc}

\usepackage{makecell}

\usepackage{graphicx}
\usepackage{subcaption} 

\usepackage{enumitem} 

\AtBeginDocument{%
  }

\setcopyright{acmlicensed}
\copyrightyear{2018}
\acmYear{2018}
\acmDOI{XXXXXXX.XXXXXXX}

\acmJournal{POMACS}
\acmVolume{37}
\acmNumber{4}
\acmArticle{111}
\acmMonth{8}

\emergencystretch=\maxdimen
\begin{document}
\settopmatter{printacmref=false}

\title{A Survey of Affective Recommender Systems: Modeling Attitudes, Emotions, and Moods for Personalization}


\author{Tonmoy Hasan}
\email{thasan1@chatlotte.edu}
\orcid{0000-0002-2116-6822}

\author{Razvan Bunescu}
\email{rbunescu@charlotte.edu}
\orcid{0000-0003-2919-3566}

\affiliation{%
  \institution{University of North Carolina at Charlotte}
  \city{Charlotte}
  \state{NC}
  \country{USA}
}

\renewcommand{\shortauthors}{Hasan and Bunescu}

\begin{abstract}
Affective Recommender Systems are an emerging class of intelligent systems that aim to enhance personalization by aligning recommendations with users’ affective states. Reflecting a growing interest, a number of surveys have been published in this area, however they lack an organizing taxonomy grounded in psychology and they often study only specific types of affective states or application domains. This survey addresses these limitations by providing a comprehensive, systematic review of affective recommender systems across diverse domains. Drawing from Scherer’s typology of affective states, we introduce a classification scheme that organizes systems into four main categories: attitude aware, emotion aware, mood aware, and hybrid. We further document affective signal extraction techniques, system architectures, and application areas, highlighting key trends, limitations, and open challenges. As future research directions, we emphasize hybrid models that leverage multiple types of affective states across different modalities, the development of large-scale affect-aware datasets, and the need to replace the folk vocabulary of affective states with a more precise terminology grounded in cognitive and social psychology. Through its systematic review of existing research and challenges, this survey aims to serve as a comprehensive reference and a useful guide for advancing academic research and industry applications in affect-driven personalization.
\end{abstract}



\maketitle

\section{Introduction and Motivation}
\label{sec:introduction}

Understanding human emotions and their impact on behavior has long been a central focus in psychology. The term "affect" was notably used in psychological studies by \citeauthor{wundt1897outline} \cite{wundt1897outline} in the late 19th century to describe the subjective experience of pleasantness or unpleasantness. Since then, the concept of affect has evolved, encompassing a wide range of emotional and psychological states that include emotions (short term), moods (longer lasting), and attitudes (stable), each impacting human behavior in unique ways \cite{ekman1992argument, scherer2000psychological}. Recognizing the powerful role of affect in human behavior, researchers in psychology and behavioral sciences have extensively studied its impact on human motivation, decision-making, and social interactions. In 1999, \citeauthor{picard1999affective} \cite{picard1999affective} proposed that by recognizing and responding to human emotions, machines could enhance human-computer interaction not only by making it more empathetic, but also by enabling systems to naturally adapt to users, improve communication, and better handle affective information, such as frustration, confusion, interest, and preference. The growing understanding of affect in human behavior laid the groundwork for the field of {\it affecting computing} \cite{picard2000affective,calvo:tac2010}, which has evolved into a broad discipline with applications to healthcare, gaming, education, and beyond. 

One of the most promising applications of affective computing is in Recommender Systems (RS), where understanding and incorporating users' affective states into recommendation algorithms can lead to significantly improved user satisfaction. Traditionally, recommender systems leverage users' explicit preferences (e.g., product ratings, stated interests) and behavioral history (e.g., clicks, browsing patterns). However, affective RS go a step further by incorporating users' emotional and psychological states, henceforth referred to as {\it affective states}, into the recommendation process. By integrating affective signals, recommender systems can dynamically adjust recommendations to align with the user’s current mood, emotional state, or long-term dispositions, leading to a more engaging and satisfying user experience \cite{Tkalčič2022}. For instance, a user’s mood can directly influence their content choices, like preferring relaxing music after a stressful day or choosing upbeat movies for a boost in mood. Similarly, a person feeling lonely may seek emotionally resonant books or supportive online communities to improve their affective state.

Integrating affective state information has shown promise in a range of recommendation domains. In music and video streaming platforms, for example, affective RS adapt content suggestions to match a user’s mood, enhancing the emotional impact of recommendations \cite{kaminskas2018music}. In education, a study by \citeauthor{Santos2014} \cite{Santos2014} explored how students' affective states, such as boredom or engagement, can be detected and used to adapt educational content accordingly. Their research demonstrated that recognizing and responding to students' emotions can enhance learning experiences and improve learning outcomes. \citeauthor{mizgajski2019affective} \cite{mizgajski2019affective} investigated how emotions influence reading choices in online news, highlighting the role of affective states in content consumption. While their study focused on news, its findings extend to e-commerce, where aligning product recommendations with users’ emotions or sentiments can enhance engagement, satisfaction, and potentially increase purchase intent \cite{pappas2017sense}. However, incorporating users' affective states into recommender systems is not an easy task. For example, emotions and moods are dynamic, context-dependent, and can vary considerably from one user to another, posing challenges in reliably detecting and responding to affective states in real time \cite{scherer2001appraisal}. Despite these challenges, affective RS have gained popularity because of their potential to significantly enhance recommendations through a nuanced understanding of users' emotional states and preferences. 

Reflecting the growing importance of leveraging user preferences and emotions in recommender systems, the number of publications on affective RS has risen significantly in recent years. 
The increasing interest calls for a comprehensive review on affective RS aimed at helping researchers and practitioners better understand their strengths, limitations, and best use cases.

\subsection{Differences between This Survey and Prior Surveys}

A number of surveys have been conducted in various subdomains of affective recommendation systems by focusing on particular types of affective states or on particular application domains. Overall, these surveys have been limited by their narrower scope and the lack of a structured approach to affective states informed by psychological theories of emotion. Furthermore, depending on their publication date, prior surveys do not cover the most recent advancements in affective RS. 
For instance, \citeauthor{katarya2016physica} \cite{katarya2016physica} provided a broad overview of affective advancements within recommender systems, but lacked a structured affective taxonomy, and did not distinguish between significantly different types of affective states such as sentiment, emotion, and mood \cite{scherer2005emotions}.  
Furthermore, their work did not consider serendipity-oriented recommender systems, even though serendipity, through its surprise component, is a complex affective state with important consequences for user satisfaction. Some affective RS surveys had a narrower scope by focusing on a specific domain. For example, \citeauthor{wang2022affective} \cite{wang2022affective} examined video applications, \citeauthor{salazar2021affective} \cite{salazar2021affective} investigated affective adaptation in education, \citeauthor{piccarra2022searching} \cite{piccarra2022searching} focused on emotion-based movie navigation, and \citeauthor{granados2021tourist} \cite{granados2021tourist} reviewed emotion recognition techniques in tourism. 
Similar to \citeauthor{katarya2016physica} \cite{katarya2016physica}, these works lacked a structured affective taxonomy and did not include serendipity-oriented recommendations. Additionally, some of them ignored long-term affective states such as mood, which play a crucial role in sustained user engagement. Other affective RS surveys focus solely on a particular type of affective state. For example, \citeauthor{al2021comprehensive} \cite{al2021comprehensive} targeted sentiment analysis techniques, categorizing works by methods such as lexicon-based or machine learning-based approaches. In the area of context-driven music recommendations, \citeauthor{kaminskas2018music} \cite{kaminskas2018music} considered only emotion-aware recommender systems. Several surveys exclusively explored serendipity in recommender systems. For example, \citeauthor{kotkov2016serendipity} \cite{kotkov2016serendipity} and \citeauthor{ziarani2021serendipity} \cite{ziarani2021serendipity} explored techniques to achieve unexpectedness and relevance, categorizing works according to the various methods used to promote serendipity, diversity, and novelty.
Works by \citeauthor{fu2023deepserendipity} \cite{fu2023deepserendipity} and \citeauthor{kaminskas2016survey} \cite{kaminskas2016survey} examined deep learning techniques and beyond-accuracy goals in serendipitous recommendations, categorizing studies by technical methods and beyond-accuracy metrics.

Notwithstanding the important contributions of these surveys, they remain fragmented and limited in scope, often focusing on particular types of affective states or on specific application domains. To the best of our knowledge, no survey has explored the full range of affective states, categorized into psychologically grounded types such as emotions, moods, or attitudes. Table~\ref{tab:survey_comparison} presents a comparative analysis of existing surveys, evaluating whether they provide a structured affective taxonomy, comprehensively cover all types of affective states, and consider diverse application domains.
\begin{table*}[t]
    \centering
    \footnotesize
    \caption{Comparison of this survey with previous work in terms of affective taxonomy, types of affective states, and application domain coverage. Application domain is \textit{Broad} if it spans diverse application areas.}
    \label{tab:survey_comparison}
    \begin{tabular}{p{4.4cm}cccccc}
        \toprule
        \multirow{2}{*}{\textbf{Survey Papers}} & \textbf{Affective} & \textbf{Attitude} & \textbf{Emotion} & \textbf{Mood} & \textbf{Serendipity} & \textbf{App.} \\
        & \textbf{taxo.} & \textbf{(stable)} & \textbf{(brief)} & \textbf{(durable)} & \textbf{(hybrid)} & \textbf{domain} \\
        \midrule
        \citeauthor{katarya2016physica} \cite{katarya2016physica} (2016) & \textcolor{purple}{$\circ$} & \textcolor{teal}{\checkmark} & \textcolor{teal}\checkmark & \textcolor{teal}\checkmark & \textcolor{purple}{$\circ$} & {\it Broad}\\
        \citeauthor{atas2021towards} \cite{atas2021towards} (2021) & \textcolor{purple}{$\circ$} & \textcolor{purple}{$\circ$} &  \textcolor{teal}\checkmark & \textcolor{purple}{$\circ$} & \textcolor{purple}{$\circ$} & {\it Broad}\\
        \citeauthor{wang2022affective} \cite{wang2022affective} (2022) & \textcolor{purple}{$\circ$} & \textcolor{teal}\checkmark & \textcolor{teal}\checkmark & \textcolor{teal}\checkmark & \textcolor{purple}{$\circ$} & Video \\
        \citeauthor{salazar2021affective} \cite{salazar2021affective} (2021) & \textcolor{purple}{$\circ$} & \textcolor{teal}\checkmark & \textcolor{teal}\checkmark & \textcolor{purple}{$\circ$} & \textcolor{purple}{$\circ$} & Education \\
        \citeauthor{piccarra2022searching} \cite{piccarra2022searching} (2022) & \textcolor{purple}{$\circ$} & \textcolor{teal}\checkmark & \textcolor{teal}\checkmark & \textcolor{teal}\checkmark & \textcolor{purple}{$\circ$} & Movie \\
        \citeauthor{granados2021tourist} \cite{granados2021tourist} (2021) & \textcolor{purple}{$\circ$} & \textcolor{teal}\checkmark & \textcolor{teal}\checkmark & \textcolor{teal}\checkmark & \textcolor{purple}{$\circ$} & Tourism \\
        \citeauthor{al2021comprehensive} \cite{al2021comprehensive} (2021) & \textcolor{purple}{$\circ$} & \textcolor{teal}\checkmark & \textcolor{purple}{$\circ$} & \textcolor{purple}{$\circ$} & \textcolor{purple}{$\circ$} & {\it Broad}\\
        \citeauthor{kaminskas2018music} \cite{kaminskas2018music} (2018) & \textcolor{purple}{$\circ$} & \textcolor{purple}{$\circ$} & \textcolor{teal}\checkmark & \textcolor{teal}\checkmark & \textcolor{purple}{$\circ$} & Music \\
        \citeauthor{kotkov2016serendipity} \cite{kotkov2016serendipity} (2016) & \textcolor{purple}{$\circ$} & \textcolor{purple}{$\circ$}  & \textcolor{purple}{$\circ$}  & \textcolor{purple}{$\circ$}  & \textcolor{teal}\checkmark & {\it Broad} \\
        \citeauthor{ziarani2021serendipity} \cite{ziarani2021serendipity} (2021) & \textcolor{purple}{$\circ$} & \textcolor{purple}{$\circ$}  & \textcolor{purple}{$\circ$}  & \textcolor{purple}{$\circ$} & \textcolor{teal}\checkmark & {\it Broad} \\
        \citeauthor{fu2023deepserendipity} \cite{fu2023deepserendipity} (2023) & \textcolor{purple}{$\circ$} & \textcolor{purple}{$\circ$} & \textcolor{purple}{$\circ$}  & \textcolor{purple}{$\circ$}  & \textcolor{teal}\checkmark & {\it Broad} \\
        \citeauthor{kaminskas2016survey} \cite{kaminskas2016survey} (2016) & \textcolor{purple}{$\circ$} & \textcolor{purple}{$\circ$}  & \textcolor{purple}{$\circ$}  & \textcolor{purple}{$\circ$}  & \textcolor{teal}\checkmark & {\it Broad}\\
        \citeauthor{fakhri2019serendipity} \cite{fakhri2019serendipity} (2019) & \textcolor{purple}{$\circ$} & \textcolor{purple}{$\circ$} & \textcolor{purple}{$\circ$} & \textcolor{purple}{$\circ$} & \textcolor{teal}\checkmark & {\it Broad} \\
        \textbf{This survey} & \textcolor{teal}\checkmark & \textcolor{teal}\checkmark & \textcolor{teal}\checkmark & \textcolor{teal}\checkmark & \textcolor{teal}\checkmark & {\it Broad} \\
        \bottomrule
    \end{tabular}
\end{table*}
As shown in Table~\ref{tab:survey_comparison}, previous surveys are limited in the types of affective states they cover or the range of applications they consider. Our survey addresses these limitations by providing a structured taxonomy that organizes affective RS across multiple application domains, offering a holistic perspective on the role of affective signals in personalized recommendations.

\subsection{Literature Search Methodology}

To conduct a comprehensive survey of affective RS, we started by systematically gathering relevant literature published between 2015 and 2025. Our approach ensured the inclusion of high-quality, relevant studies that address various types of affective states, such as sentiment, emotion, and mood, across all application domains within recommender systems. We performed an initial search across major academic databases, including SpringerLink, IEEE Xplore, ACM Digital Library, ScienceDirect, Web of Science, and Wiley. The search query was designed to capture studies that explore affective aspects in recommender systems. Specifically, we used boolean operators to construct the following search string in article titles:
\begin{itemize}[label={}]
    \item {\tt (recommend OR recommender OR recommendations OR recommendation) AND (sentiment OR unexpectedness OR unexpected OR emotion OR emotional OR affective OR mood OR empathetic OR emotions OR surprise OR surprising OR serendipity OR serendipitous OR psychology)}.
\end{itemize}
Besides journal articles, we ensured the inclusion of papers from peer-reviewed conferences known for publishing impactful work in recommender systems and related fields, such as RecSys, SIGIR, CIKM, AAAI, NeurIPS, WWW, WSDM, KDD, UMAP, EMNLP, ACL, NAACL, and ICML.

The initial search yielded approximately 1,500 articles. We removed duplicate records across the databases and then we screened the remaining papers by examining titles and abstracts to assess their relevance to our study’s focus on affective RS. Factors considered included primary research focus, contribution relevance, and publication venue prominence. After this preliminary screening, we retained 172 articles. To identify additional relevant works outside of the initial target timeframe, we applied a snowball sampling technique, based on the studies cited within the initially selected papers. This approach added 35 more papers that fell outside the designated publication timeframe, selected based on their relevance, citation impact, and the significance of their publication venues. This approach ensured that our database included influential works that align with the study’s focus on affective RS. The final corpus consisted of 207 papers, covering a broad range of topics within sentiment, emotion, mood, and serendipity aware recommender systems.

\subsection{Contributions of This Survey}

The primary aim of this survey is to provide a comprehensive review of affective RS, covering different types affective states, including sentiment, emotion, and mood, across diverse application domains. The key contributions of this survey are as follows:

\begin{itemize}
    \item We introduce a novel classification scheme for affective recommender systems based on Scherer’s typology of affective states \cite{scherer2000psychological}, categorizing publications according to the distinct types of affective states that they target (Section~\ref{sec:classfication_scheme}). 
    \item  We systematically review how attitude (Section~\ref{sec:attitude-rs}), emotion (Section~\ref{sec:emotion-rs}), mood (Section~\ref{sec:mood-rs}), and hybrid affective states (Section~\ref{sec:hybrid-rs}) are modeled, detected, and integrated into recommendation strategies (Section~\ref{sec:affective_recsys}) across different applications domains (Section~\ref{sec:dataset_application}). 
    \item We explore the current challenges and gaps in affective recommender systems, offering suggestions for future research directions, in particular encouraging the development of datasets and recommender models that target a broader range of affective states (Section~\ref{sec:future}).
\end{itemize}

\section{Background and Classification Scheme} 
\label{sec:classfication_scheme}

\begin{figure}[t]
    \centering
    \includegraphics[width=\linewidth]{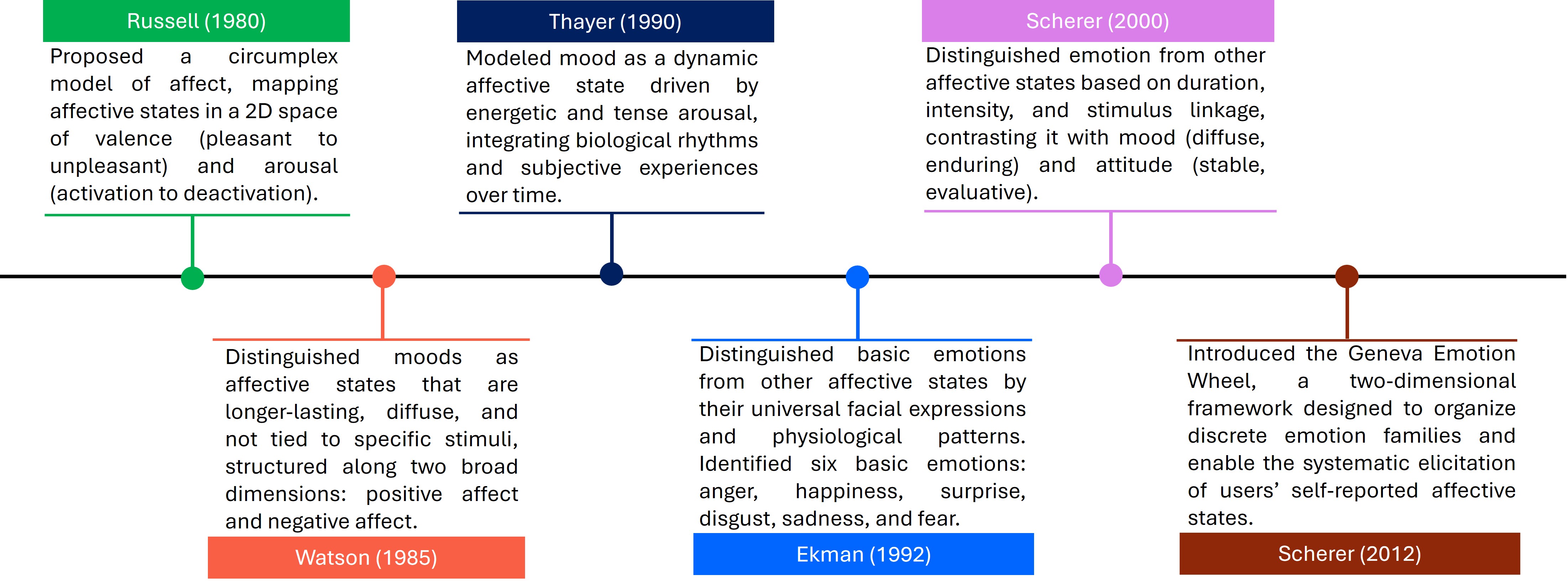}
    \caption{Summaries of foundational psychological theories of affect.}
    \label{fig:affect_theory}
\end{figure}

Affective computing is a multidisciplinary field that uses computational methods to study and develop systems that recognize, interpret, process, and simulate the various affective aspects of human behavior \cite{picard2000affective,tao2005affective}. Over the years, numerous studies in psychology, social sciences, and philosophy have tried to define and characterize different types of affective states that people experience, and resolve the inherent fuzziness of the various language categories and folk terms that people use. For example,  even though the concept of "emotion" is used very frequently, the question of "{\it what is an emotion}" rarely obtains the same answer from different individuals \cite{scherer2000psychological}. The peripheral theory of William James \cite{james1922emotions}, for instance, conflates the terms "emotion" and "feeling". However, according to the component process model of Scherer, the term "feeling" should be reserved only for the subjective experience component of emotion, distinct from facial and vocal expression (motor), action tendencies (motivational), bodily symptoms (neurophysiological), or appraisal (cognitive) components. As part of the effort to clarify what emotions are, \citeauthor{watson1985toward} \cite{watson1985toward} focused on classifying moods as affective states distinct from emotions, proposing a model in which affective states are arranged on positive and negative affect dimensions. According to their research, moods are diffuse, longer-lasting states that subtly influence perception and behavior without being directly tied to specific stimuli. \citeauthor{thayer1990biopsychology} \cite{thayer1990biopsychology} explored mood as a specific type of affective state that is distinct from emotions and attitudes, providing a model that integrates mood with physiological arousal levels. \citeauthor{ekman1992argument} \cite{ekman1992argument} distinguished basic emotions from other affective states, such as moods and attitudes, by identifying the unique characteristics of emotions (e.g., facial expressions and physiological patterns) that may not manifest in other types of affective states. \citeauthor{russell1999core} \cite{russell1999core} introduced core affect as a foundational concept that captures a broad spectrum of affective states beyond discrete emotions, characterizing them using continuous dimensions such as valence (pleasantness vs. unpleasantness) and arousal (activation vs. deactivation). \citeauthor{barrett2006solving} \cite{barrett2006solving} discussed how affective states, such as mood and emotion, arise from a core affect and highlighted the continuum of affective experiences that range from general affective feelings to specific emotional responses. Psychologist \citeauthor{scherer2005emotions} \cite{scherer2000psychological, scherer2005emotions} distinguished emotion from other types of affective states, emphasizing that affective states vary in their duration, intensity, and the degree to which they are linked to specific stimuli. Scherer's typology includes affective attitudes as stable evaluative dispositions, which are less intense than emotions, relatively enduring, and reflect affectively colored beliefs towards objects or individuals. Examples of attitudes include preferences, liking, loving, hating, valuing, and desiring. Emotions represent more intense, short-lived responses to specific stimuli, often accompanied by physiological and mental changes. Emotions encompass a wide range of states, such as anger, sadness, joy, fear, shame, pride, elation, desperation, and surprise. Moods are more diffuse and enduring affective states that subtly influence perceptions and behaviors over extended periods, with examples such as cheerful, gloomy, irritable, listless, depressed, and buoyant. Informed by Scherer's typology of affective states, the Geneva Emotion Wheel \cite{sacharin2012geneva} was developed as a two dimensional representation of discrete emotion families, in order to enable the elicitation of affective descriptions from users in a standard, systematic way. Figure~\ref{fig:affect_theory} summarizes some of the foundational psychological theories of affect discussed above.

The diverse nature of affective experiences, ranging from attitudes and moods to emotions, has inspired various computational frameworks for personalization in RS. When designing affective RS, it is important to consider underlying psychological theories of affective states.  Building upon these theories, this survey proposes a novel classification scheme that mirrors the psychological categorization of affective states into attitudes, emotions, moods, and hybrid states. Additionally, the survey characterizes how each of these affective dimensions influences user preferences and impacts recommender system outcomes.

\subsection{Classification Scheme for Affective Recommender Systems}

The classification of affective states has been explored from multiple perspectives in psychology and affective computing, with contributions from scholars such as Watson, Thayer, Ekman, Russell, Barrett, and Scherer. Categorical models (e.g., Ekman’s six basic emotions), dimensional models (e.g., Russell’s circumplex model, Thayer’s arousal-tension model), and core affect theories (e.g., Barrett’s conceptual act theory), while providing useful affective frameworks, they primarily focus on emotions and often do not systematically differentiate emotions from other affective states such as attitudes and moods. Comparatively, Scherer’s typology offers a more comprehensive framework by distinguishing affective states based on their duration, intensity, and cognitive involvement. Unlike categorical and dimensional models, which primarily emphasize emotions, Scherer’s classification incorporates a broader spectrum of affective phenomena, including stable evaluative dispositions, i.e., attitudes, transient affective states, i.e., emotions, and longer-lasting states, i.e., moods. This distinction is particularly relevant to affective RS, where different types of affective states influence recommendations in different ways: emotions capture immediate user responses, moods shape long-term consumption patterns, and attitudes reflect stable preferences.

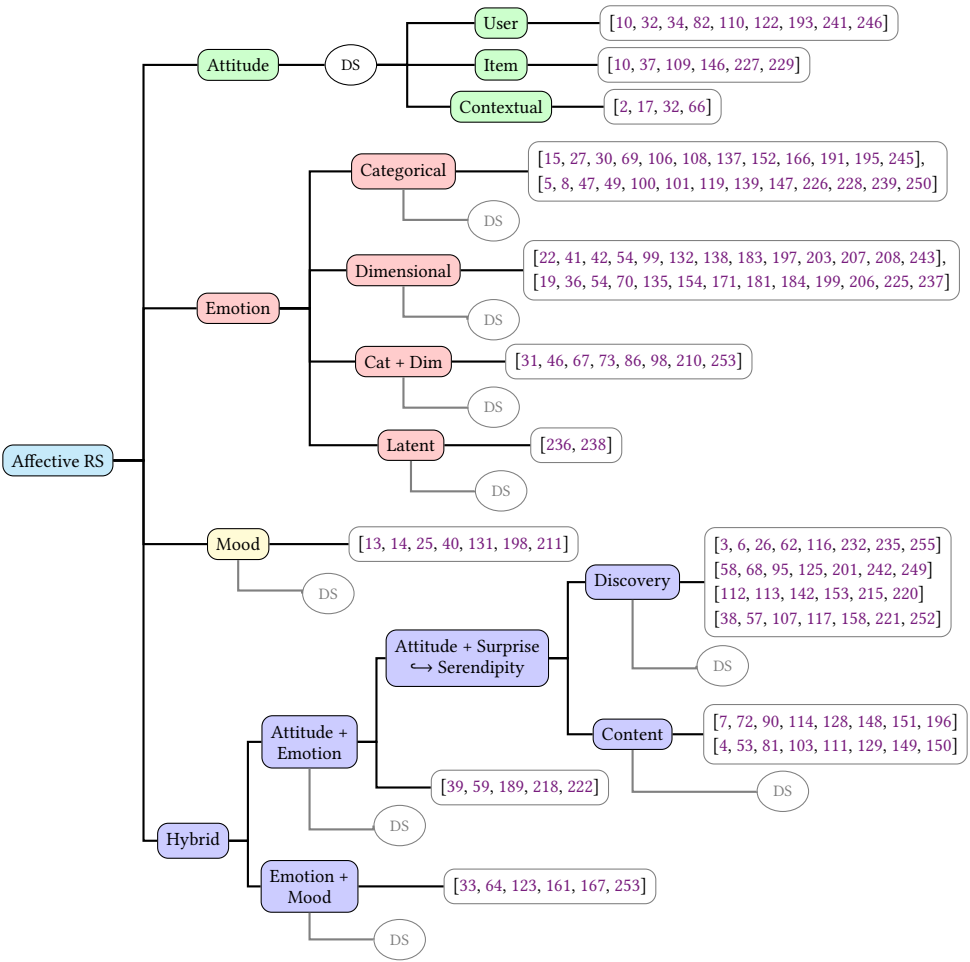
\begin{figure}[ht]
\centering
\begin{tikzpicture}[
  box/.style={circle, rectangle, draw, rounded corners, minimum width=0.05cm, minimum height=0.4cm,  minimum size=0.05cm, align=center, font=\scriptsize, fill=#1!20},
  line/.style={draw, thick},
]

\node[box=cyan] (root) at (0,0) {Affective RS};

\node[box=green]   (att)    at ($(root)+(2.4,5.2)$) {Attitude};
\node[box=red]     (emo)    at ($(root)+(2.4, 2.0)$) {Emotion};
\node[box=yellow]    (mood)   at ($(root)+(2.4,-1.1)$) {Mood};
\node[box=blue]  (hybrid) at ($(root)+(1.8,-5.0)$) {Hybrid};

\node[draw, ellipse, minimum size=0.05cm, font=\tiny] (att_ds) at ($(att)+(1.5,0)$) {DS};

\node[box=green] (user)    at ($(att_ds)+(2,0.55)$) {User};
\node[box=green] (item)    at ($(att_ds)+(2,0)$)   {Item};
\node[box=green] (context) at ($(att_ds)+(2.0,-0.55)$) {Contextual};

\node[draw=gray, box=green, fill=none, minimum height=0.2cm] (cite_user)  at ($(user)+(3.3,0)$)   {\cite{li2016moviesentiment, cai2019topicsentiment, liu2021deepsentiment, alam2022newssetiment, cai2022deepsentiment, zhang2015phrasesentiment, shi2022sengr, zhang2021sifn, hyun2018scalablesentiment}};
\node[draw=gray, box=green, fill=none, minimum height=0.2cm] (cite_item)  at ($(item)+(2.7,0)$)   {\cite{chen2019evalsentiment, li2020timesentiment, n2023coursesentiment, wu2024bayesiansentiment, wu2020sentirec, alam2022newssetiment}};
\node[draw=gray, box=green, fill=none, minimum height=0.2cm] (cite_context)  at ($(context)+(2.15,0)$)   {\cite{abbasi2021tourismsentiment, asani2021restaurantsentiment, cai2019topicsentiment, 10360257}};

\node[box=red] (cat)     at ($(emo)+(2.2,1.8)$)  {Categorical};
\node[box=red] (dim)     at ($(emo)+(2.2,0.5)$)  {Dimensional};
\node[box=red] (catdim)  at ($(emo)+(2.2,-0.7)$) {Cat + Dim};
\node[box=red] (latent)     at ($(emo)+(2.3,-1.8)$) {Latent};

\node[draw=gray, box=green, fill=none, minimum height=0.2cm] (cite_cat)  at ($(cat)+(4.45,0)$)   {\shortstack[l]{
\cite{meng2018exploitemotion, BREITFUSS2021715, ARAMANDA2023120190, poirson2019customeremotion, zhang2024empathetic, kuo2005musicemotion, Santos02012016, guo2019collaborative, su2020emotioncloth, lopez2021edurecomsys, shen2020peia, leung2020textemotion},\\
\cite{moscati2024musicemotion, deng2015musicemotion, adru2024musicemotion, s21061997, 7930506, 7902195, LIM2017179, 10.1145/3151759.3151817, 10.1145/2964284.2964327, 10.1145/2072298.2072043, 10.1145/3426020.3426119, 10.1145/3573834.3574478, dodd2022facialemotion}}};
\node[draw=gray, box=green, fill=none, minimum height=0.2cm] (cite_dim)  at ($(dim)+(4.5,0)$) {\shortstack[l]{\cite{zhang2024affectivevideo, KIM2018214, costa2013emotion, tkalvcivc2010using, 10.3389/fpsyg.2019.00675, mizgajski2019affective, CHHEDA2023383, https://doi.org/10.1155/2022/9548486, 6378304, 6901531, ferrato2022meta4rs, tkalcic2018emotiontutorial, benini2011connotative},\\
\cite{ayata2018musicemotion, gonzalez2019deepemotion, 10720053, tkalcic2013affectiveimages, yoon2012musicemotion, han2024musicemotion, ferrato2022meta4rs, revathy2023lyemobert, 10.1145/2503385.2503484, wen2022fairemotion, mehrabian1996pleasure, canini2013connotative, 10.1145/2700171.2791042}}};
\node[draw=gray, box=green, fill=none, minimum height=0.2cm] (cite_catdim)  at ($(catdim)+(3,0)$)   {\cite{Zheng2016, kim2024indoor, tripathi2019emoware, ishanka2017prefiltering, https://doi.org/10.1111/bjet.13209, deng2015emotional, gonzalez2006embedding, hasan2025leveraging}};
\node[draw=gray, box=green, fill=none, minimum height=0.2cm] (cite_latent)  at ($(latent)+(2.2,0)$)   {\cite{yousefian2021emotion, yin2024collaborative}};

\node[draw=gray, text=gray, ellipse, minimum size=0.05cm, font=\tiny] (cat_ds)  at ($(cat)+(1.2,-0.65)$)   {DS};
\node[draw=gray, text=gray, ellipse, minimum size=0.05cm, font=\tiny]  (dim_ds)  at ($(dim)+(1.2,-0.65)$)   {DS};
\node[draw=gray, text=gray, ellipse, minimum size=0.05cm, font=\tiny] (catdim_ds)  at ($(catdim)+(1.2,-0.6)$)   {DS};
\node[draw=gray, text=gray, ellipse, minimum size=0.05cm, font=\tiny] (latent_ds)  at ($(latent)+(1.2,-0.6)$)   {DS};

\node[draw=gray, box=green, fill=none, minimum height=0.2cm] (cite_mood)  at ($(mood)+(3,0)$)   {\cite{tang2021improve, ueda2016recipe, marshall2016mood, andjelkovic2019moodplay, andjelkovic2016moodplay, chen2015mood, bontempelli2022flowmoods}};
\node[draw=gray, text=gray, ellipse, minimum size=0.1cm, font=\tiny] (mood_ds)  at ($(mood)+(1.2,-0.65)$)   {DS};

\node[box=blue] (att_emo)   at ($(hybrid)+(1.55,1.3)$)  {Attitude +\\ Emotion};
\node[box=blue] (emo_mood)  at ($(hybrid)+(1.55,-0.6)$) {Emotion +\\ Mood};

\node[draw=gray, box=green, fill=none, minimum height=0.2cm] (cite_emo_mood)  at ($(emo_mood)+(3.2,0)$)   {\cite{LIU2023103256, polignano2021emotion, piazza2017affective, cai2007emotion, Zheng2016, gilda2017musicmoodemotion}};
\node[draw=gray, text=gray, ellipse, minimum size=0.1cm, font=\tiny] (emo_mood_ds)  at ($(emo_mood)+(1.2,-0.7)$)   {DS};

\node[draw=gray, text=gray, ellipse, minimum size=0.1cm, font=\tiny] (att_emo_ds)  at ($(att_emo)+(1.2,-1.1)$) {DS};

\node[box=blue] (serendip)  at ($(att_emo)+(2.1,1.1)$)   {Attitude + Surprise\\$\hookrightarrow$ Serendipity};
\node[draw=gray, box=green, fill=none, minimum height=0.2cm] (cite_att_emo)  at ($(att_emo)+(2.8,-0.6)$)   {\cite{https://doi.org/10.1155/2022/7246802, 10.1145/3627043.3659561, chen2018emotional, sertkan2022emotion, wang2023airline}};

\node[box=blue] (discovery) at ($(serendip)+(2.2,1.0)$) {Discovery};
\node[draw=gray, box=green, fill=none, minimum height=0.2cm] (cite_discovery)  at ($(discovery)+(2.6,0)$)   {\shortstack[l]{
\cite{yang2017improving, ge2020serendipity, ZIARANI2021115660, boo2023session, afridi2018serendipity, xu2020serendipity, li2019serendipity, adamopoulos2014unexpectedness}\\
\cite{grange2019little, lu2012serendipitous, fu2023serendipity, taramigkou2013escape, kawamae2010serendipitous, zhang2021next, zhang2012serendipity}\\
\cite{li2020unexpectedness, li2020unexpected, onuma2009surprise, murakami2007metrics, wang2023item, 8392508}\\
\cite{fu2024serendipity, chen2019serendipity, zheng2015unexpectedness, wang2020impacts, pandey2018recommending, li2020directional, lee2020serendipity}}};
\node[draw=gray, text=gray, ellipse, minimum size=0.1cm, font=\tiny] (discovery_ds)  at ($(discovery)+(1.2,-1.1)$)   {DS};

\node[box=blue] (content) at ($(serendip)+(2.2,-1)$) {Content};
\node[draw=gray, box=green, fill=none, minimum height=0.2cm] (cite_content)  at ($(content)+(2.7,0)$)   {\shortstack[l]{
\cite{sugiyama2015towards, afridi2020facilitating, niu2021luckyfind, maake2019serendipitous, niu2018adaptive, jenders2015serendipity, hasan2023serendipity, li2024variety}\\
\cite{li2020utility, fan2018implementing, niu2018surprise, huang2018learning, kotkov2020does, niu2017framework, maccatrozzo2017sirup, aditya2025engineering}}};
\node[draw=gray, text=gray, ellipse, minimum size=0.1cm, font=\tiny] (content_ds)  at ($(content)+(2,-0.75)$)   {DS};

\draw[line] (root.east) -- ++(0.4,0) |- (att.west);
\draw[line] (root.east) -- ++(0.4,0) |- (emo.west);
\draw[line] (root.east) -- ++(0.4,0) |- (mood.west);
\draw[line] (root.east) -- ++(0.4,0) |- (hybrid.west);

\draw[line] (att.east) -- ++(0.4,0) |- (att_ds.west);
\draw[line] (att_ds.east) -- ++(0.4,0) |- (user.west);
\draw[line] (att_ds.east) -- ++(0.4,0) |- (item.west);
\draw[line] (att_ds.east) -- ++(0.4,0) |- (context.west);
\draw[line] (user.east) -- ++(0.4,0) |- (cite_user.west);
\draw[line] (item.east) -- ++(0.4,0) |- (cite_item.west);
\draw[line] (context.east) -- ++(0.4,0) |- (cite_context.west);

\draw[line] (emo.east) -- ++(0.4,0) |- (cat.west);
\draw[line] (emo.east) -- ++(0.4,0) |- (dim.west);
\draw[line] (emo.east) -- ++(0.4,0) |- (catdim.west);
\draw[line] (emo.east) -- ++(0.4,0) |- (latent.west);

\draw[line] (cat.east) -- ++(0.4,0) |- (cite_cat.west);
\draw[line] (dim.east) -- ++(0.4,0) |- (cite_dim.west);
\draw[line] (catdim.east) -- ++(0.4,0) |- (cite_catdim.west);
\draw[line] (latent.east) -- ++(0.4,0) |- (cite_latent.west);

\draw[draw=gray, line] (cat.south) -- ++(0,-0.4) -| (cat_ds.west);
\draw[draw=gray, line] (dim.south) -- ++(0,-0.4) -| (dim_ds.west);
\draw[draw=gray, line] (catdim.south) -- ++(0,-0.4) -| (catdim_ds.west);
\draw[draw=gray, line] (latent.south) -- ++(0,-0.4) -| (latent_ds.west);

\draw[line] (mood.east) -- ++(0.4,0) |- (cite_mood.west);
\draw[draw=gray, line] (mood.south) -- ++(0,-0.4) -| (mood_ds.west);

\draw[line] (hybrid.east) -- ++(0.25,0) |- (att_emo.west);
\draw[line] (hybrid.east) -- ++(0.25,0) |- (emo_mood.west);

\draw[line] (emo_mood.east) -- ++(0.4,0) |- (cite_emo_mood.west);
\draw[draw=gray, line] (emo_mood.south) -- ++(0,-0.35) -| (emo_mood_ds.west);

\draw[draw=gray, line] (att_emo.south) -- ++(0,-0.8) -| (att_emo_ds.west);

\draw[line] (att_emo.east) -- ++(0.25,0) |- (serendip.west);
\draw[line] (att_emo.east) -- ++(0.25,0) |- (cite_att_emo.west);
\draw[line] (serendip.east) -- ++(0.25,0) |- (discovery.west);
\draw[line] (serendip.east) -- ++(0.25,0) |- (content.west);

\draw[line] (content.east) -- ++(0.25,0) |- (cite_content.west);
\draw[draw=gray, line] (content.south) -- ++(0,-0.55) -| (content_ds.west);

\draw[line] (discovery.east) -- ++(0.25,0) |- (cite_discovery.west);
\draw[draw=gray, line] (discovery.south) -- ++(0,-0.9) -| (discovery_ds.west);

\end{tikzpicture}
\caption{Classification scheme for organizing affective recommender systems. DS represents the data sources used as input in recommender systems, and they include user, item, and contextual information.}
\label{fig:affective}
\end{figure}

Drawing from Scherer’s categorization of affective states, this survey classifies affective RS into four main categories: attitude, emotion, mood, and hybrid. Each category is further subdivided based on methodological approaches, as illustrated in Fig.~\ref{fig:affective}. This classification provides a structured foundation for organizing affect-aware recommendation techniques, enabling a systematic comparison of methodologies across various domains.
\begin{itemize}

    \item {\bf Attitude:} This category includes sentiment-aware recommender systems that focus on users' stable evaluative judgments, such as positive or negative sentiments toward items. Notably, all recommender systems inherently incorporate some form of user attitude, either implicitly through historical interaction data (e.g., clicks, purchases, ratings) or explicitly through task definitions that aim to predict whether a user will like an item. However, in this survey, we focus specifically on those RS that go beyond such notions of liking or disliking an item and instead aim to extract and model attitudes from unstructured, user-generated content such as textual reviews, social media posts and comments, blog writing, or forum discussions.
    
    \item {\bf Emotion:} This category covers emotion-aware recommender systems, which model transient affective responses arising from specific stimuli. To capture users’ emotional states, these systems typically leverage categorical models (e.g., Ekman’s six basic emotions) and dimensional models (e.g., Russell’s valence-arousal framework, Plutchik's multidimensional framework).
    \item {\bf Mood:} This category includes mood-aware recommender systems focusing on longer lasting affective states, such as cheerfulness, melancholy, or relaxation. Unlike emotions, moods persist over extended periods and influence long-term content consumption patterns. These systems leverage behavioral signals, contextual information, and interaction histories to infer users' mood-driven preferences.
    \item {\bf Hybrid:} This category refers to recommender systems that integrate multiple types of affective states for improved personalization. Examples include attitude-emotion hybrid approaches, which combine users' evaluative preferences (e.g., sentiment) with transient affective states (emotion), and emotion-mood hybrid systems, which account for both emotional responses and mood states to enhance content adaptation.
    \begin{itemize}
        \item {\bf Serendipity}: A special subcategory of hybrid systems is represented by serendipity-oriented recommender systems. While serendipity is not itself a basic affective state, it can be equated as an {\it unexpected} and {\it pleasant} experience. This characterization reflects a combination of two types of affective components: (i) the emotion of unexpectedness or surprise, and (ii) the positive evaluative attitude toward the outcome.
    \end{itemize}
\end{itemize}

\section{General Architecture of Affective Recommender Systems}
\label{sec:affective_recsys}

\begin{figure}[t]
    \centering
    \includegraphics[width=1\linewidth]{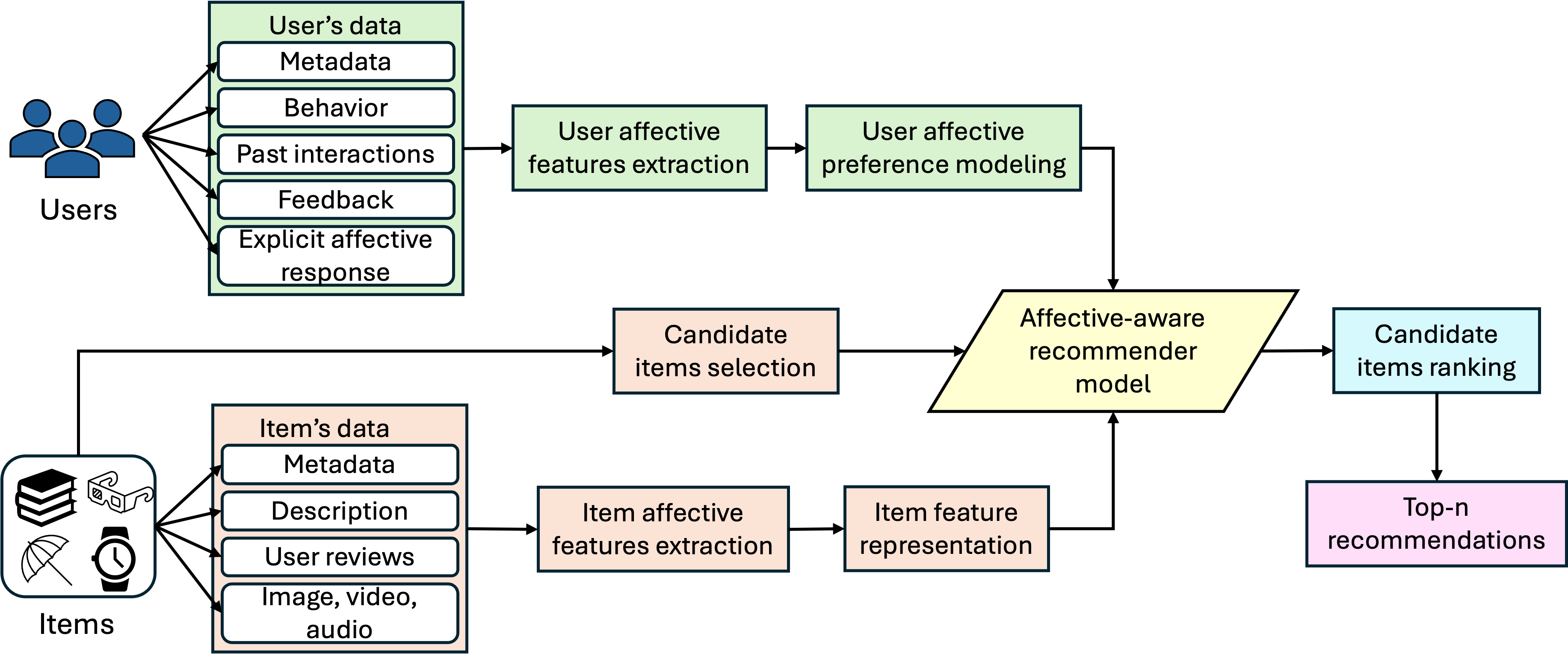}
    \caption{A general architecture of affective recommender systems.}
    \label{fig:generic_affectivers}
\end{figure}

Traditional recommender systems primarily rely on explicit ratings, past interactions, and user metadata to predict preferences \cite{zhang2019recsys, burke2002hybrid, bobadilla2013recsys, lu2015recsys}. However, these methods overlook the affective dimension of user experience, which plays a crucial role in shaping content preferences. Affective RS bridge this gap by integrating users’ emotions, moods, and attitudes into the recommendation process, ensuring that the system adapts not only to what users like, but also to how they feel during different stages of interaction~\cite{tkalcic2011affective}, such as at the point of entry, during content consumption, or upon exit. Figure~\ref{fig:generic_affectivers} illustrates a general framework for affective RS consisting of three key stages of affective modeling and integration, as described below:
\begin{enumerate}[leftmargin=15pt]
    \item {\bf User Affective Modeling.} Understanding a user’s affective states requires integrating diverse data sources. These include user metadata (e.g., age, location), behavioral signals (e.g., watch duration, clicks, browsing history), past interactions (e.g., purchases, likes, reviews), feedback (e.g., ratings, textual comments), and explicit affective input (e.g., self-reported emotions). Affect-aware systems perform feature extraction to convert these signals into structured affective representations. For instance, in e-commerce, textual affective features can be derived from reviews or social media feedback, capturing emotional reactions to previous purchases. In music and video streaming platforms, audio-visual cues, such as facial expressions, offer deeper insight into affective preferences. Once extracted, these features are used to construct a dynamic model of the user’s affective profile, capturing users' attitude, emotional states and mood. This modeling can be rule-based, or driven by machine learning and deep learning techniques that continuously adapt based on user feedback. For example, in movie recommendation, the system might suggest nostalgic films when the user feels lonely or action films during heightened excitement.
    \item {\bf Item Affective Feature Representation.} This component focuses on extracting affect-relevant attributes from item content. Inputs may include item metadata (e.g., genre, brand), descriptive information (e.g., specifications), user-generated content (e.g., reviews), and multimedia signals (e.g., trailers, product images, music). These features are processed and structured using feature engineering and embedding methods. For instance, affective information from textual content can be extracted through feature engineering approaches such as part-of-speech tagging, syntactic parsing, and affective lexicon matching. Embedding methods, such as Word2Vec, GloVe, and BERT embedding, are then used to transform this information into numerical representations suitable for downstream modeling. In image- and video-based recommender systems, convolutional neural networks (CNNs) are commonly employed to extract visual-emotional cues, such as color tone, brightness, and facial expressions. These affective features are typically embedded into lower-dimensional spaces for integration with user modeling.
    \item {\bf Affect-Aware Recommendation Generation.} In this final stage, user affective and item affective features representations are combined to generate personalized recommendations. Candidate items are first retrieved using standard techniques such as collaborative filtering, content-based filtering, or hybrid methods. Affective signals are then used to re-rank these candidates, prioritizing items that align with the user’s current affective states or long-term tendencies. In addition to affective cues, some systems further incorporate contextual information, such as time of day, location, or weather conditions, to refine personalization, thereby enhancing user engagement, satisfaction, and emotional resonance.
\end{enumerate}
In the following sections, we survey Affective RS approaches according to the four major categories of our classification scheme: attitude-aware (Section~\ref{sec:attitude-rs}), emotion-aware (Section~\ref{sec:emotion-rs}), mood-aware (Section~\ref{sec:mood-rs}), and hybrid systems (Section~\ref{sec:hybrid-rs}). 

\section{Attitude-Aware Recommender Systems}
\label{sec:attitude-rs}

Attitude-aware recommender systems aim to incorporate users' evaluative preferences, often inferred from affective expressions of sentiment or opinion, into the recommendation process. These preferences reflect relatively stable user attitudes toward items, attributes, or content categories. For instance, in an e-commerce setting, the following review illustrates how users convey nuanced evaluations that extend beyond simple like or dislike.

\begin{center}
\begin{tcolorbox} [width=0.95\textwidth,
    colback=white,
    colframe=gray!10!black, 
    colbacktitle=black!60, 
    coltitle=white,        
    title=Example Review,
    fonttitle=\bfseries,
    boxrule=0.5pt,
    left=4pt,
    right=4pt,
    top=2pt,
    bottom=2pt,
    enhanced,
    sharp corners,
    before skip=6pt,
    after skip=6pt]
\small
``\textit{I love how lightweight and modern this laptop is, but the battery life is disappointing}''
\end{tcolorbox}
\end{center}
From this review, a system can infer that the user values sleek and portable designs but is dissatisfied with poor battery performance. In future interactions, such cues can guide the system to prioritize laptops with long battery life and filter out bulkier models with limited endurance. This example highlights how even a single review can reveal multiple attitudinal signals, which, when effectively captured, can substantially improve recommendation relevance.

Due to sustained progress in sentiment analysis techniques over more than two decades of research, attitude-aware RS approaches vastly outnumber the other types of affective RS, as such they have received significantly more attention in literature reviews. As a quick-reference resource, Table~\ref{tab:attitude} lists representative sentiment-aware recommender system papers along with their targeted sentiment fusion strategies, application domains, attitude modeling approaches, and data sources.  

To infer user attitudes, affective RS rely on a variety of data sources, including textual reviews, explicit ratings, behavioral signals such as clicks or watch time, and even demographic information. These systems use affective modeling through feature engineering, machine learning, or deep learning to extract evaluative cues that shape user-item representations, influence item rankings, or adjust predicted ratings. Understanding both the sources and nature of affective data is therefore essential for building effective attitude-aware recommender systems. Affective signals can originate from {\it user behavior} (Section~\ref{sec:user}) and {\it item data} (Section~\ref{sec:item}), whereas their impact on recommendations can be modulated by {\it contextual conditions} (Section~\ref{sec:contextual}).

\subsection{User-Centered Attitude Data}
\label{sec:user}

User-centered data is foundational for modeling attitudinal preferences in affective RS. This category includes user-generated content such as reviews, microblogs, comments, past interactions, ratings, browsing history, clicks, and demographic metadata. These user-related signals are frequently leveraged to extract affective information and shape sentiment-aware recommendation models utilizing approaches ranging from lexicon based approaches to deep learning and language model based approaches, as summarized below:
\begin{itemize}
    \item {\bf Lexicon Based Approaches}. These methods rely on predefined dictionaries of sentiment-laden words, such as SentiWordNet, VADER, AFINN, and TextBlob, to incorporate user-related information into the recommendation process. Aspect-based sentiment analysis (ABSA) is often employed with lexicon-based methods to extract sentiment at a feature level. For instance, \citeauthor{10.1145/3109859.3109905} \cite{10.1145/3109859.3109905} and A2SPR \cite{10.1145/3414841} employed SentiWordNet for aspect-level sentiment scoring within collaborative filtering and graph-based models, respectively, utilizing user reviews, comments, and check-in history to enhance recommendation quality by aligning with user preferences. Similarly, Serrano-\citeauthor{SERRANOGUERRERO2024122340} \cite{SERRANOGUERRERO2024122340} applied SenticNet and VADER in a fuzzy linguistic healthcare recommender, using user reviews and comments to improve hospital service rankings through sentiment analysis. \citeauthor{li2020timesentiment} \cite{li2020timesentiment} also relied on user reviews and comments, extending ABSA by incorporating temporal dynamics into SentiWordNet-based sentiment scoring to capture users’ evolving preferences over time.
    \item {\bf Traditional Machine Learning Approaches}. Traditional machine learning methods have been widely employed to classify sentiment and incorporate it into recommendation systems utilizing user information. For instance, \citeauthor{li2016moviesentiment} \cite{li2016moviesentiment} combined rule-based sentiment extraction using LingPipe with SVM and logistic regression to mine user sentiment from microblog posts, comments, and movie-watching histories for predicting movie preferences. Building on the idea of leveraging user-generated content, \citeauthor{cai2019topicsentiment} \cite{cai2019topicsentiment} utilized social media posts and user interactions (e.g., likes) to extract sentiment features, which were then integrated into a matrix factorization model (SIO-TMF) enhanced with LDA for improved friend recommendation in social networks. Extending sentiment-based modeling to address data sparsity issues, \citeauthor{zhang2015phrasesentiment} \cite{zhang2015phrasesentiment} proposed a matrix factorization framework that incorporates sentiment-enhanced feature-opinion pairs extracted from user reviews, effectively mitigating cold-start challenges where explicit user ratings are sparse or unavailable.
    \item {\bf Deep Learning Approaches}. Deep learning introduced greater accuracy in sentiment-aware recommendation systems. \citeauthor{n2023coursesentiment} \cite{n2023coursesentiment} employed Elman Recurrent Neural Networks (ERNN) to perform sentiment classification on user course reviews extracted from tweets and comments, integrating sentiment signals into an e-learning recommender system that outperformed traditional models. \citeauthor{liu2021deepsentiment} \cite{liu2021deepsentiment} earlier proposed a multilingual sentiment-aware model that uses bidirectional GRUs with attention mechanisms to perform aspect-based sentiment analysis on multilingual user reviews and ratings, enhancing rating prediction across languages. To improve both scalability and rating prediction accuracy, \citeauthor{hyun2018scalablesentiment} \cite{hyun2018scalablesentiment} introduced SentiRec, a two-step CNN-based recommendation model that encodes user sentiment from reviews into fixed-size vectors for user and item representations.
    \item {\bf Language Model Based Approaches}. The emergence of With the emergence of large pre-trained language models such as BERT, sentiment-aware recommender systems have significantly improved their ability to capture contextualized user preferences from user reviews, enabling more accurate and personalized recommendations across various domains. A range of sentiment-aware recommendation systems have leveraged BERT to enrich user modeling, including \citeauthor{zhang2021sifn} \cite{zhang2021sifn}, who fused sentiment signals in a review-encoding network to alleviate sparsity and cold-start problems; \citeauthor{wu2024bayesiansentiment} \cite{wu2024bayesiansentiment}, who combined BERT with Bayesian Personalized Ranking to jointly capture semantic and visual sentiment representations for improved predictive accuracy; \citeauthor{ZHANG2022871} \cite{ZHANG2022871}, who incorporated topic-aware sentiment classification to enhance recommendation diversity; and \citeauthor{alam2022newssetiment} \cite{alam2022newssetiment}, who analyzed affect-laden news content and revealed that user demographics influence affective interpretation.
    \end{itemize}
In summary, user rooted affective modeling in sentiment aware recommender systems encompasses a wide spectrum of methods, from lexicon based sentiment scoring and traditional classifiers to neural architectures and language model based techniques. These approaches leverage the richness of user related data to enhance personalization, interpretability, and the overall effectiveness of recommendation systems.

\subsection{Item-Centered Attitude Data}
\label{sec:item}

Item-centered data provides important targets of attitude signals, especially when combined with aspect oriented sentiment expressed in reviews. This category includes structured product metadata (e.g., specifications, categories, and attributes), item descriptions, or item-relevant data from social media. A variety of approaches integrate these signals into item modeling to enhance recommendation quality:
\begin{itemize}
    \item {\bf Item Features}. Explicit product attributes and descriptive metadata are often aligned with sentiment signals to construct affect-aware item representations. \citeauthor{chen2019evalsentiment} \cite{chen2019evalsentiment} combined product specifications (e.g., price, RAM, battery life) with sentiment scores derived from user reviews using WordNet-based similarity and SentiWordNet polarity. These signals were integrated into a utility-based ranking function to generate interpretable, sentiment-aware recommendations. Similarly, \citeauthor{li2020timesentiment} \cite{li2020timesentiment} aligned aspect metadata from item descriptions (e.g., CPU, RAM) with sentiment scores from reviews to build aspect-level sentiment matrices. These were incorporated into a similarity based matrix factorization model where item-item similarity was influenced by both latent and sentiment-aware aspect signals.
    \item {\bf Social Media}. Sentiment-oriented text from external sources such as social media can inform item representations. For instance, \citeauthor{n2023coursesentiment} \cite{n2023coursesentiment} constructed item representations for e-learning courses by leveraging unstructured textual data sourced from social networking platforms such as Twitter. A hybrid sentiment analysis framework, comprising ITF-IDF, Word2Vec, Hybrid N-gram features, and an Elman Recurrent Neural Network (ERNN) was employed to extract sentiment polarities from the course-related texts. The resulting sentiment profiles were used to represent each course and subsequently matched with user sentiment vectors using multiple similarity measures to generate personalized recommendations.
    \item {\bf Multimodal Data}. Multimodal signals enhance item representations and improve recommendation performance. For instance, \citeauthor{wu2024bayesiansentiment} \cite{wu2024bayesiansentiment} combined visual sentiment from movie posters with semantic sentiment from user reviews. These signals were embedded into a four-dimensional tensor encompassing users, items, visual sentiment, and semantic sentiment, with tensor decomposition applied to learn latent factors for ranking via Bayesian Personalized Ranking (BPR).
\end{itemize}
Taken together, these approaches highlight how item-rooted affective signals, ranging from structured specifications and aspect-level sentiment to multimodal inputs and unstructured social discourse, enable nuanced modeling of item affective preferences. Beyond supporting personalized content retrieval, item sentiment representations also contribute to explanation generation and fairness assessment, underscoring their broader utility in enhancing both the performance and transparency of affect-aware recommender systems.

\subsection{Contextual Data}
\label{sec:contextual}

Contextual data provides essential situational awareness in affective RS, ensuring that recommendations are not only sentiment-aligned but also feasible and temporally appropriate. This category includes information such as location, time, weather, and other environmental or situational variables that influence the practical relevance of recommendations at the time they are delivered. In the domain of tourism recommendation, \citeauthor{abbasi2021tourismsentiment} \cite{abbasi2021tourismsentiment} proposed a sentiment-aware system that integrated contextual information, specifically, location, time, and weather, with user check-ins, ratings, and review texts. User sentiment preferences were then refined through contextual filtering: only attractions located in the user’s current city, operating at the time of recommendation, and appropriate to the prevailing weather conditions (e.g., prioritizing indoor venues during rain or snow) were considered, ensuring that recommended items were both sentiment-aligned and contextually feasible. A similar approach was applied in restaurant recommendation \cite{asani2021restaurantsentiment}, using location and operating hours as contextual filters. Extending beyond static context, \citeauthor{cai2019topicsentiment} \cite{cai2019topicsentiment} segmented user histories over time and applied matrix factorization within each segment to capture time-based changes in sentiment, enhancing adaptability in friend recommendations.

Overall, by incorporating contextual factors like location, time, and weather, the system can filter out impractical suggestions, making sure the recommendations are not only attitude-aligned, but also usable and contextually appropriate. \\

{\scriptsize
\begin{longtable}{p{0.85cm}p{4.6cm}p{2.5cm}p{2.55cm}p{1.5cm}}
  \caption{{\bf Attitude-aware} recommender systems papers. {\footnotesize\textit{Note:} In the Attitude (Att.) Modeling column, \textit{discrete} refers to predefined sentiment categories (e.g., positive, strong positive, negative, strong negative, neutral); the number in parentheses indicates how many sentiment levels were used. \textit{Continuous} refers to sentiment polarity scores, and \textit{implicit} refers to latent or learned sentiment representations. Abbreviations: Att. = Attitude, Ra = Ratings, Rev = Reviews, Des = Description, MF = Matrix Factorization, FM = Factorization Machine, NFM = Neural Factorization Machine, DeepFM = Deep Factorization Machine, AFM = Attentional Factorization Machine, Cos = Cosine Similarity, JS = Jaccard Similarity, LR = Logistic Regression, SVM = Support Vector Machine, SVMR = Support Vector Machine Regression, Attn = Attention, SGD = Stochastic Gradient Descent, KL-div = Kullback–Leibler Divergence, NARRE = Neural Attentional Rating Regression, DAML = Dual Attention Memory Learning, LDA = Latent Dirichlet Allocation, USE = Universal Sentence Encoder, GCN = Graph Convolutional Network, KGCN = Knowledge Graph Convolutional Network, RF = Random Forest, DT = Decision Tree, NB = Naive Bayes, CF = Collaborative Filtering, MLP = Multi-Layer Perceptron, ERNN = Elman Recurrent Neural Network, POI = Point of Interest.}}
  \label{tab:attitude}\\
  \toprule
  \textbf{Papers} & \textbf{Data Source} & \textbf{Application} & \textbf{Techniques} & \textbf{Att. Modeling} \\
  \midrule
  \endfirsthead
  \toprule
  \textbf{Papers} & \textbf{Data Source} & \textbf{Application} & \textbf{Techniques} & \textbf{Att. Modeling} \\
  \midrule
  \endhead
  \multicolumn{5}{r}{\textit{Continued on next page}} \\
  \midrule
  \endfoot


  \bottomrule
  \endlastfoot
    \cite{li2016moviesentiment} & Ra, item metadata, interactions & Movie, TV Program & LR, SVMR & Discrete (5)\\
    \midrule
    \cite{cai2019topicsentiment} & Blog posts, interactions, time, likes, comment  & Social Media & MF, Cos & Discrete (2) \\
    \midrule
    \cite{zhang2015phrasesentiment} & Rev, Ra & E-commerce & Lexicon-based approach & Discrete (2) \\
    \midrule
    \cite{n2023coursesentiment} & Tweets, comments, course description & Education & ERNN, Cos, JS & Discrete (3) \\
    \midrule
    \cite{hyun2018scalablesentiment} & Rev, Ra & E-commerce & CNN, DeepCoNN, D-Attn &  Continuous\\
    \midrule
    \cite{cai2022deepsentiment} & Rev, Ra & E-commerce & LDA, DNN, MF, FM, Cos & Discrete (2) \\
    \midrule
    \cite{shi2022sengr} & Rev, Ra & E-commerce, restaurant & GCN, Co-Attn, FM & Implicit \\
    \midrule
    \cite{wu2020sentirec} & Clicks, news title & News & Transformers & Discrete (2) \\
    \midrule
    \cite{zhang2021sifn} & Rev, Ra, User ID, Item ID & E-commerce & BERT, Attn & Discrete (3) \\
    \midrule
    \cite{ZHANG2022871} & Rev & Movie & BERT, LDA, TF-IDF & Discrete 3 \\
    \midrule
    \cite{10.1145/3397271.3401330} & Rev, interactions  & E-commerce, restaurant & LSTM, RNN, Attn & Discrete (3) \\
    \midrule
    \cite{https://doi.org/10.1002/cpe.6359} & Rev, Ra  & E-commerce & TF-IDF & Discrete (3) \\
    \midrule
    \cite{chen2019evalsentiment} & Rev, item specifications & E-commerce & SentiWordNet & Continuous \\
    \midrule
    \cite{alam2022newssetiment} & Clicks, age, gender, date, news metadata & News & BERT, RippleNet, TF-IDF & Continuous  \\
    \midrule
    \cite{10.1145/3404835.3462943, 10.1145/3511808.3557558} & Rev, Ra & E-commerce & DeepCoNN, MF, NARRE & Discrete (2)  \\
    \midrule
    \cite{li2020timesentiment} & Rev, Ra, item metadata & E-commerce & SentiWordNet, LDA, MF & Discrete (3) \\
    \midrule
    \cite{ghosal-etal-2019-deepsentipeer} & Research papers, peer Rev & Education & CNN, MLP, VADER, USE & Discrete (3) \\
    \midrule
    \cite{9964412, 9377188} & Rev, Ra, drug names & Healthcare & KGCN / LR, NB, DT, RF & Discrete (3)  \\
    \midrule
    \cite{10.1145/3488560.3498515} & Rev, Ra, item metadata & E-commerce & Reinforcement learning & Discrete (3) \\
    \midrule
    \cite{10.1145/2442810.2442816} & News titles, URLs, publication dates, location & News & LDA, SVM & Discrete (3) \\
    \midrule
    \cite{8978583} & Rev, Ra, user metadata, item specification & E-commerce & fastText & Discrete (5) \\
    \midrule
    \cite{10047910} & Rev, Ra & E-commerce & BERT, LDA, MF, RNN & Implicit \\
    \midrule
    \cite{10.5555/3172077.3172271} & Rev, Ra & Movie, restaurant & Latent Factor Model & Implicit \\
    \midrule
    \cite{10.1145/3097983.3098122} & User ID, item ID, location, Rev, item metadata & Tourism & LDA & Discrete (2) \\
    \midrule
    \cite{https://doi.org/10.1155/2022/4940401} & Rev, Ra, item metadata & Restaurants & MF & Discrete (2) \\
    \midrule
    \cite{10.1145/2600428.2609579} & Rev, Ra & Restaurants, phones & K-means, MF, rule-based & Discrete (2) \\
    \midrule
    \cite{10431746} & Rev, Ra, item metadata & Education & Bi-LSTM, Attn, Word2Vec & Discrete (5) \\
    \midrule
    \cite{8675268} & Rev, Ra & E-commerce, restaurant & LSTM, FM, Attn & Implicit \\
    \midrule
    \cite{10360257} & Check-ins, User comments, POI metadata & POI & LSTM, GCN, Attn & Discrete (2) \\
    \midrule
    \cite{10509543} & User ID, item ID, Rev, Ra & E-commerce & LSTM, CNN, TextBlob & Continuous \\
    \midrule
    \cite{8445585} & User posts, user-item metadata & Healthcare & CNN, RNN, LSTM & Implicit \\
    \midrule
    \cite{10.1145/3320435.3320457} & Rev, Ra, item metadata & Movie, book & Rule-based & Discrete (2) \\
    \midrule
    \cite{abbasi2021tourismsentiment} & User check-in, Rev, Ra, location, time, weather & Tourism & SentiWordNet & Discrete (2) \\
    \midrule
    \cite{asani2021restaurantsentiment} & Rev, check-in, location, time & Restaurant & SentiWordNet, Cos & Discrete (2) \\
    \midrule
    \cite{9103168} & Ra, tweets, movie metadata & Movie & VADER & Discrete (3) \\
    \midrule
    \cite{10.1145/3414841} & Rev, Ra, item metadata & E-commerce & SentiWordNet, Cos & Discrete (3) \\
    \midrule
    \cite{9006442} & Tweets, time, POI metadata & Tourism & VADER, TextBlob & Discrete (3) \\
    \midrule
    \cite{artemenko2020using} & Rev, location & Tourism & Rule-based & Discrete (5) \\
    \midrule
    \cite{https://doi.org/10.1155/2020/8892552} & Tweets, user metadata & Healthcare & Rule-based & Discrete (3) \\
    \midrule
    \cite{10.1145/3298689.3347024} & Rev, Ra, item metadata & Movie & Stanford CoreNLP, KL-div & Discrete (5) \\
    \midrule
    \cite{liu2021deepsentiment} & Rev, Ra & E-commerce & Bi-GRU, CNN, Attn & Implicit \\
    \midrule
    \cite{10047910} & Rev, Ra & E-commerce & Bi-RNN, BERT, LDA, MF & Implicit \\
    \midrule
    \cite{li-etal-2021-recommend-reason} & Rev, Ra & E-commerce & CNN, BERT & Implicit \\
    \midrule
    \cite{9149913} & Rev, Ra, item titles & E-commerce & BERT, FM, Transformer & Discrete (3) \\
    \midrule
    \cite{10.1371/journal.pone.0248695} & Rev, Ra & E-commerce & CF, Cos & Discrete (3) \\
    \midrule
    \cite{10.1145/3109859.3109905} & Rev, Ra & Hotels, E-commerce & MF, SGD & Implicit \\
    \midrule
    \cite{8784145} & Rev, Ra, location & Tourism & NB, SVM & Discrete (3) \\
    \midrule
    \cite{GARCIACUMBRERAS20136758} & Rev, Ra, movie metadata & Movie & KNN, SVM, MF & Discrete (2) \\
    \midrule
    \cite{8796367} & Rev, place description and images & Tourism & SentiWordNet, LDA & Continuous \\
    \midrule
    \cite{10.1145/3184558.3191583} & Tweets, location, time & Location & SVM & Continuous \\
    \midrule
    \cite{10.1145/2481492.2481505} & Check-in, Rev, location & Location & MF, SentiWordNet & Continuous \\
    \midrule
    \cite{https://doi.org/10.1111/exsy.12991} & Rev, Ra, interactions, item metadata & Movie & KNN, Stanford CoreNLP & Continuous \\
    \midrule
    \cite{lu2021revcore} & Past interaction, Rev, movie title \& metadata & Movie & Transformer, GNN & Discrete (2) \\
    \midrule
    \cite{darraz2025integrated} & Ra, Rev, user-item metadata, past interaction & Restaurant, hotel & BERT & Discrete (2)  \\
\end{longtable}
}
\normalsize

\section{Emotion-Aware Recommender Systems}
\label{sec:emotion-rs}

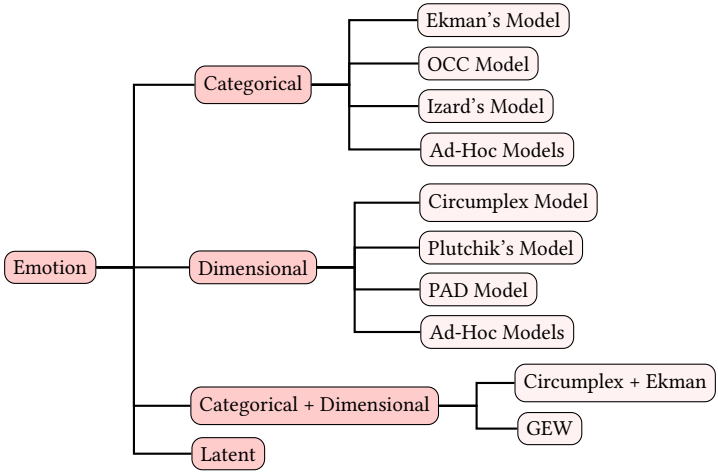
\begin{figure}[t]
\centering
\begin{tikzpicture}[
  box/.style={circle, rectangle, draw, rounded corners, minimum width=0.05cm, minimum height=0.4cm,  minimum size=0.05cm, align=center, font=\footnotesize, fill=#1!20},
  line/.style={draw, thick},
]

\node[box=red] (emotion) at (0,0) {Emotion};

\node[box=red] (cat)     at ($(emotion)+(2.7,2.4)$) {Categorical};
\node[box=red] (dim)     at ($(emotion)+(2.7,0)$) {Dimensional};
\node[box=red] (catdim)  at ($(emotion)+(3.52,-1.82)$) {Categorical + Dimensional};
\node[box=red] (latent)   at ($(emotion)+(2.37,-2.45)$) {Latent};

\node[box=pink] (ekman)   at ($(cat)+(3.2,0.85)$) {Ekman's Model};
\node[box=pink] (occ)     at ($(cat)+(3,0.28)$) {OCC Model};
\node[box=pink] (izard)   at ($(cat)+(3.1,-0.28)$) {Izard's Model};
\node[box=pink] (adhoc_cat) at ($(cat)+(3.25,-0.85)$) {Ad-Hoc Models};

\node[box=pink] (circum)  at ($(dim)+(3.4,0.85)$) {Circumplex Model};
\node[box=pink] (plutchik) at ($(dim)+(3.3,0.26)$) {Plutchik's Model};
\node[box=pink] (pad)     at ($(dim)+(3,-0.29)$) {PAD Model};
\node[box=pink] (adhoc_dim) at ($(dim)+(3.25,-0.85)$) {Ad-Hoc Models};

\node[box=pink] (circum_ekman) at ($(catdim)+(4,0.3)$) {Circumplex + Ekman};
\node[box=pink] (gew)          at ($(catdim)+(3.13,-0.3)$) {GEW};

\draw[line] (emotion.east) -- ++(0.5,0) |- (cat.west);
\draw[line] (emotion.east) -- ++(0.5,0) |- (dim.west);
\draw[line] (emotion.east) -- ++(0.5,0) |- (catdim.west);
\draw[line] (emotion.east) -- ++(0.5,0) |- (latent.west);

\draw[line] (cat.east) -- ++(0.5,0) |- (ekman.west);
\draw[line] (cat.east) -- ++(0.5,0) |- (occ.west);
\draw[line] (cat.east) -- ++(0.5,0) |- (izard.west);
\draw[line] (cat.east) -- ++(0.5,0) |- (adhoc_cat.west);

\draw[line] (dim.east) -- ++(0.5,0) |- (circum.west);
\draw[line] (dim.east) -- ++(0.5,0) |- (plutchik.west);
\draw[line] (dim.east) -- ++(0.5,0) |- (pad.west);
\draw[line] (dim.east) -- ++(0.5,0) |- (adhoc_dim.west);

\draw[line] (catdim.east) -- ++(0.5,0) |- (circum_ekman.west);
\draw[line] (catdim.east) -- ++(0.5,0) |- (gew.west);

\end{tikzpicture}
\caption{Taxonomy for emotion-aware recommender systems. Representative papers for each of the 4 emotion representation models are listed in Table~\ref{tab:emotion}.}
\label{fig:emotion-taxonomy}
\end{figure}

Adapting to users' emotional states at the various stages of the recommendation process (entry, consumption, and exit) can significantly improve recommendation relevance, user satisfaction, and engagement \cite{tkalcic2011affective, zheng2013role}. For instance, emotion recognition at the entry stage can help recommend calming content after a stressful event, or amusing content during moments of sadness. The importance of modeling emotions lies in how they reflect short-term, immediate user states that traditional approaches may overlook.  To represent emotions, two types of models have been widely adopted in emotion-aware recommender systems:
\begin{itemize}
    \item \textbf{Categorical} models, which define emotions as discrete states (Section~\ref{subsub:categorical}).
    \item \textbf{Dimensional} models, which describe emotions along continuous scales such as valence and arousal (Section~\ref{subsub:dimensional}).
\end{itemize}
In Figure~\ref{fig:emotion-taxonomy} we shows the taxonomy used to organize emotion-aware recommender systems. Table~\ref{tab:emotion} provides a concise tabular overview of a comprehensive set of emotion-aware RS papers, specifying emotion modeling approaches, application domains, and data sources. In the rest of this section, we analyze representative work, focusing on their emotion modeling strategies and how these are applied across different domains.

\subsection{Categorical Models of Emotion}

\label{subsub:categorical}
\begin{figure}[t]
    \centering
    \begin{subfigure}[b]{0.3\textwidth} 
        \centering
        \includegraphics[width=\linewidth]{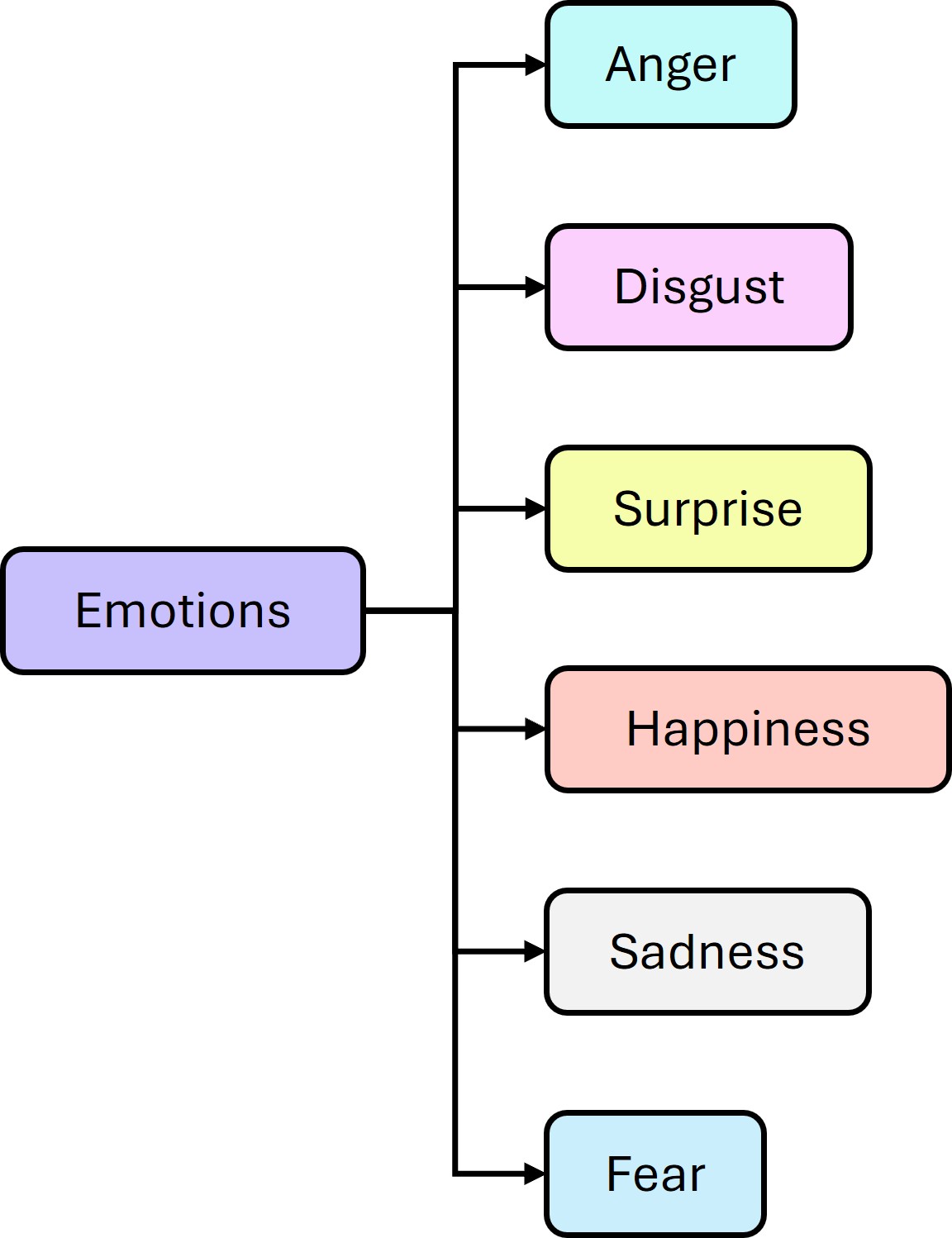} 
        \caption{Ekman's six basic emotions}
        \label{fig:categorical}
    \end{subfigure} 
    \begin{subfigure}[b]{0.62\textwidth}
        \centering
        \includegraphics[width=\linewidth]{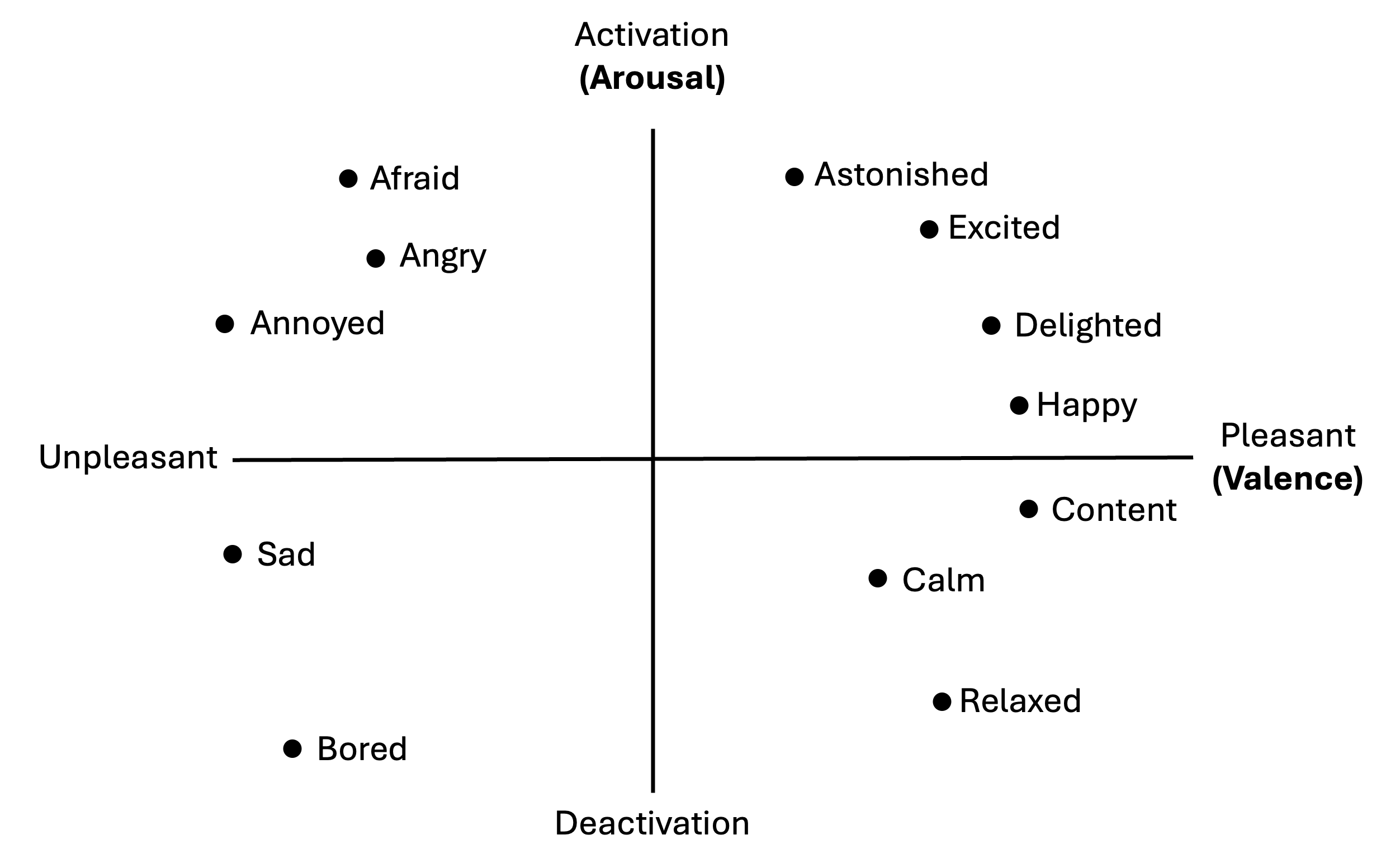} 
        \caption{Russell's dimensional model}
        \label{fig:dimensional}
    \end{subfigure}
    
    \caption{Representative categorical (a) and dimensional (b) emotion models.}
    \label{fig:combined}
\end{figure}

Categorical models of emotion have their roots in the field of psychology, where researchers have long sought to classify emotions into discrete categories. One of the most influential contributions in this area came from \citeauthor{ekman1971universals} \cite{ekman1971universals, ekman1992argument}, who identified six basic emotions: {\it joy}, {\it sadness}, {\it anger}, {\it fear}, {\it disgust}, and {\it surprise} (Figure~\ref{fig:categorical}). These emotions were shown to be universally recognized across cultures, making them a cornerstone of emotion research. The Ortony, Clore, and Collins (OCC) model of \citeauthor{ortony1988cognitive} \cite{ortony1988cognitive} defines 22 discrete emotions and two cognitive states based on appraisals of events, actions, and objects. \citeauthor{izard2013human}'s~\cite{izard2013human} Differential Emotions Theory posits that ten discrete emotions are fundamental, innate components of human experience, each characterized by unique neural and expressive patterns. Apart from the above categorical frameworks, various affective RS approaches have developed their own, ad-hoc set of emotion categories based on their suitability for a particular application domain. 

Categorical models of emotion have provided a strong theoretical basis for incorporating emotions into recommender systems. By mapping user behavior and content attributes to discrete emotional states, such as joy or sadness, systems can deliver personalized content that aligns with the user’s current affective state in an interpretable manner. Below we summarize illustrative affective RS approaches from  each of the four major classes:
\begin{itemize}
    \item {\bf Ekman's Model}. \citeauthor{leung2020textemotion}\cite{leung2020textemotion} classified emotions into happiness, sadness, anger, fear, disgust, surprise, and neutral using the Tweets Affective Classifier (TAC), a deep learning architecture featuring bidirectional LSTM and CNN layers, in order to adapt movie recommendations. \citeauthor{dodd2022facialemotion}\cite{dodd2022facialemotion} employed facial emotion recognition (FER) to detect happiness, sadness, anger, and other emotions from microexpressions, enabling emotionally adaptive systems. In another notable work, \citeauthor{deng2015musicemotion} \cite{deng2015musicemotion} extracted emotional contexts from user microblogs, categorizing them into 2D, 7D, and 21D vectors linked to music preferences within specific time windows. Emotion-aware methods such as User-based Collaborative Filtering with Emotion (UCFE) incorporated emotional similarity into the recommendation process \cite{guo2019collaborative,10.1145/2964284.2964327}.
    \item {\bf OCC Model}. \citeauthor{moshfeghi2011handling} \cite{moshfeghi2011handling} applied an OCC-based extractor to movie reviews and plot summaries, creating 22-dimensional binary vectors representing the five most frequent emotions per movie. User vectors were formed by summing emotion vectors of previously rated movies, weighted by rating scores, enabling affective similarity computation in collaborative filtering.
    \item {\bf Izard’s Model}. \citeauthor{poirson2019customeremotion}~\cite{poirson2019customeremotion} collected emotion-oriented data by asking users to rate films on eight emotional attributes adapted from Izard’s Differential Emotions Scale (DES), alongside overall preference ratings. These emotion ratings were then used to compute user-to-user similarity, enabling the system to predict preferences for unrated items based on emotionally similar users.
    \item {\bf Ad-Hoc Categorical Models}. In film music recommendation, the Music Affinity Graph (MAG) \cite{kuo2005musicemotion} associated musical features like melody and rhythm with predefined emotion categories, including 15 groups such as joy, sadness, fear, anger, and gratitude. Random walk mechanisms refined these associations, ensuring emotionally aligned recommendations by penalizing features tied to unrelated emotions. The MIRROR (eMotIon on Reviews for RecOmmendeRsystems) framework by \citeauthor{meng2018exploitemotion} \cite{meng2018exploitemotion} integrated positive and negative emotions from user reviews using various emotion categories, incorporating emotional influence through global weighting and local regularization to enhance personalization. In conversational systems, \citeauthor{zhang2024empathetic} \cite{zhang2024empathetic} considered nine emotion types (e.g., happy, negative, and surprise), fusing emotion signals with knowledge graphs to enable emotion-aware item recommendation and emotion-aligned response generation. 
\end{itemize}

\subsection{Dimensional Models of Emotion}  
\label{subsub:dimensional}

Dimensional models offer researcher an alternative framework for understanding affective states by emphasizing continuous representations rather than discrete categories. One of the most widely adopted dimensional theories is \citeauthor{russell1980circumplex}’s Circumplex Model \cite{russell1980circumplex, russell1985multidimensional}, which organizes emotions along two primary dimensions: valence (pleasure-displeasure) and arousal (activation-deactivation) (Figure~\ref{fig:dimensional}). This bidimensional space effectively captures the hedonic tone and energy levels associated with emotional states, allowing researchers to map a wide range of emotions onto a continuous plane. For instance, joy is characterized by high valence and high arousal, while sadness corresponds to low valence and low arousal. \citeauthor{plutchik1980emotion} \cite{plutchik1980emotion} introduced a psychoevolutionary dimensional framework that organizes emotions in a circumplex structure. This model identifies eight primary bipolar emotions, such as joy versus sadness and anger versus fear, arranged in a three-dimensional space reflecting emotional intensity, polarity, and similarity. Unlike Russell’s model, Plutchik’s structure does not explicitly rely on valence and arousal but instead emphasizes the evolutionary and functional relationships among emotions, including their potential to blend into complex states. Another influential dimensional framework is the Pleasure-Arousal-Dominance (PAD) model by \citeauthor{mehrabian1974approach} \cite{mehrabian1974approach}, developed in the context of environmental psychology. The PAD model characterizes emotional responses along three continuous dimensions: pleasure, arousal, and dominance. Finally, while most dimensional approaches in affective recommender systems draw from established psychological theories, a subset of studies that we named Ad-Hoc adopt domain-specific or semantically constructed affective spaces.

Building on this theoretical foundation, dimensional models have found extensive applications in emotion-aware recommender systems, enabling dynamic tracking of users' affective trajectories. Unlike categorical approaches, which classify emotions into discrete labels, dimensional models facilitate a more fluid and context-sensitive representation of emotional states, which enables recommender systems to model gradual transitions, such as shifts from excitement to relaxation or from calmness to agitation. Below we summarize illustrative affective RS approaches from each of the four major dimensional categories:
\begin{itemize}
     \item {\bf Circumplex Model}. \citeauthor{revathy2023lyemobert} \cite{revathy2023lyemobert} applied Russell’s Circumplex framework to map song lyrics into discrete emotion categories (e.g., happy, sad, angry, relaxed) by partitioning valence-arousal values and leveraging fine-tuned BERT embeddings to extract emotionally salient features, thereby improving music recommendation accuracy through a combination of semantic similarity and emotional mapping.  Similarly, other recommender systems have adopted the valence-arousal model for real-time emotion detection using physiological signals such as Galvanic Skin Response (GSR) and Electrocardiogram (ECG) \cite{ayata2018musicemotion}, with CNN-based models trained on these inputs achieving superior predictive performance over traditional classifiers \cite{gonzalez2019deepemotion}. Beyond conventional domains, this dimensional framework has also been applied to museum recommendations, where users’ valence-arousal states, combined with location tracking, were used to dynamically personalize visitor itineraries \cite{ferrato2022meta4rs}.
     \item {\bf Plutchik's Model}. Plutchik’s emotion theory inspired the Emotion-aware Transformer (EmoTER) for generating robust and fair explanations in recommender systems \cite{wen2022fairemotion}. EmoTER embedded emotional features using the NRC Emotion Lexicon and incorporated multi-task learning to align recommendations and explanations with users’ affective states. 
     \citeauthor{10.1145/2700171.2791042} \cite{10.1145/2700171.2791042} utilized Plutchik’s eight basic emotions to generate emotion vectors from YouTube comments by mapping words through the NRC (National Research Council of Canada) Emotion Lexicon. These vectors are then integrated into a context-aware recommender system using collaborative ranking to deliver personalized film recommendations based on users’ emotional preferences.
     \item {\bf PAD Model}. This framework has been widely adapted in affective computing, often using the term Valence-Arousal-Dominance (VAD), where "Valence" replaces "Pleasure." Several recommender systems incorporate this model to align content with user emotions. In image recommendation, users’ emotions are inferred from facial expressions and mapped into the valence-arousal-dominance (VAD) space to generate affective classes, which are then used to enrich user profiles in a content-based recommender system for more emotionally aligned recommendations \cite{tkalcic2013affectiveimages}.  The Hangul font recommendation system from \cite{KIM2018214} extends the PAD model to PADO (adding an Organized–Free dimension), using crowdsourced evaluations to map fonts and user preferences into emotional space, enabling personalized font selection through distance-based matching.
     \item {\bf Ad-Hoc Dimensional Models}. \citeauthor{benini2011connotative} \cite{benini2011connotative} proposed a movie recommender based on a semantic connotative space with three axes: natural (warm vs. cold), temporal (dynamic vs. slow), and energetic (impactful vs. minimal), derived from bipolar adjective pairs, reflecting stylistic and affective dimensions not explicitly tied to valence or arousal. \citeauthor{canini2013connotative} \cite{canini2013connotative} extended this work by predicting scene coordinates using audiovisual and film features from movie scenes via Support Vector Regression (SVR), removing the need for manual annotation. 
 \end{itemize}
Despite the widespread adoption of dimensional models, not all affective RS approaches accurately characterize their affective representations. For example, the study from \cite{yoon2012musicemotion} mischaracterized Thayer’s model \cite{thayer1990biopsychology} as a valence-arousal model; however, while the valence-arousal model is a well-established approach for emotion representation, Thayer’s model instead describes mood states along energy and tension axes, without directly categorizing affect in terms of positive or negative valence or arousal levels. This misrepresentation reflects a broader issue in affective computing, namely the conflation of mood and emotion models, which can lead to conceptual inconsistencies in emotion-aware recommendation design.

Overall, dimensional models offer a powerful foundation for emotion-aware recommender systems, enabling more fluid, context-sensitive, and personalized recommendations across diverse applications. Their adaptability across textual, physiological, and visual modalities provides a promising avenue for future developments in affective computing-driven personalization.


\subsection{Categorical + Dimensional Models of Emotion}

While categorical and dimensional models are often treated as distinct paradigms in affective computing, several works have explored hybrid approaches in affective RS that integrate the strengths of both, namely the interpretability of discrete emotion categories and the granularity of continuous affective dimensions.  
\begin{itemize}
    \item {\bf Circumplex + Ekman}. EmoWare \cite{tripathi2019emoware} integrates both categorical and dimensional emotion models by first positioning emotions within the 2D valence-arousal circumplex, a dimensional framework. From this space, it identifies emotional states like joy, fear, and sadness in different quadrants to define the emotional character of videos. These three discrete emotions, inspired by Ekman’s basic emotion theory, are then used as categorical labels to annotate content and track user responses, enabling a hybrid affective recommendation process.

    \item {\bf Geneva Emotion Wheel}. The Geneva Emotion Wheel (GEW) \cite{sacharin2012geneva} consists of 20 discrete emotion terms arranged around a circle and positioned such that the horizontal dimension indicates valence (negative to positive), while the vertical dimension indicates control or power (low to high).
    \citeauthor{kim2024indoor} \cite{kim2024indoor} leveraged GEW in their  Emotion-oriented Recommender System for Indoor Environmental Quality (ERS-IEQ), where the 20 emotion categories defined by GEW, such as joy, anger, fear, and interest, were inferred from multimodal data including facial expressions, voice, and physiological signals. A user's emotional state is then encoded as a point in polar coordinates (R, $\theta$), combining categorical emotion types ($\theta$) with dimensional intensity levels (R). 
\end{itemize}
Other affective recommender systems have similarly adopted hybrid approaches, combining dimensional scores with discrete emotion labels to enrich affective representations of users and items. For instance, in music recommendation, \citeauthor{deng2015emotional} \cite{deng2015emotional} developed a hybrid system using a three dimensional framework Resonance-Arousal-Valence (RAV) and categorical emotion representations inspired by the OCC model. Acoustic features are processed via SVR and variational Bayesian models to predict RAV vectors, with emotion labels used for training. \citeauthor{gonzalez2006embedding} \cite{gonzalez2006embedding} developed a Smart Prediction Assistant (SPA) that recommends educational courses in an e-learning marketplace. SPA embeds emotional context via valence scores and categorical traits (e.g., hopeful, shy) derived from questionnaires, behavior, and demographics, and then pairs each course with a short, attribute-focused text tailored to the user’s dominant emotional traits. 



\subsection{Latent Representation of Emotion}

Recent advances in emotion-aware recommender systems have explored latent representation of emotion, wherein affective cues are inferred from latent user behaviors rather than labeled emotion data. These approaches do not assign explicit emotion categories (e.g., joy or sadness) or map emotions onto predefined dimensions (e.g., valence, arousal). Instead, they extract emotion-inspired signals from patterns such as interaction histories, content preferences, or behavioral traits, and encode them into latent representations used within recommendation models. For example, \citeauthor{yousefian2021emotion} \cite{yousefian2021emotion} represent a user’s emotion state as a four dimensional Exponential Moving Average (EMA) vector over keyboard and mouse features: keystroke count, mean key hold time, mouse click count, and mean mouse button hold time. When a user selects a track, the contemporaneous EMA vector is stored with the track identifier together with an implicit one to five rating inferred from click, play, and download events, yielding a user-music log that pairs items with the user’s state at interaction time. Recommendations are then produced by comparing the current EMA vector to stored vectors using adjusted cosine similarity and a weighted sum predictor over neighbors, making the EMA vector the sole emotion representation throughout retrieval and ranking. \citeauthor{yin2024collaborative} \cite{yin2024collaborative} introduced Emotion-aware Implicit Matrix Factorization (EIMF), which integrates explicit ratings with latent “emotional” signals derived from implicit behaviors (clicks, views, purchases). Emotion is represented in low-dimensional user and item embeddings learned by the Implicit Matrix Factorization (IMF) component from one-hot behavior data, capturing global association patterns rather than psychologically defined emotions. The explicit ratings are modeled by the Emotion-aware Matrix Factorization (EMF) component, which maps users and items into high-dimensional embeddings whose inner products represent the explicit user-item evaluation captured from the rating data. These two embedding sets are fused via linear and nonlinear layers to generate final recommendation scores for movies, books, music, and images.

Overall, latent emotion representations offer a practical alternative to the development of emotion-aware recommender systems in real world settings where explicit emotion labels are scarce. \\

{\scriptsize
\begin{longtable}{p{0.4cm}p{0.65cm}p{3.2cm}p{1.3cm}p{2.75cm}p{3.2cm}} 
  \caption{{\bf Emotion-aware} recommender systems papers. {\footnotesize\textit{Note:} Cat. = Categorical, Dim. = Dimensional, Rev = Review, Ra = Rating, Cos = Cosine Similarity, MMG = Mixed Media Graph, GA = Genetic Algorithm, PC = Pearson Correlation, RCNN = Recurrent Convolutional Neural Network, LGBMClassifier = Light Gradient Boosting Machine Classifier, NMF = Non-negative Matrix Factorization, VAE = Variational Autoencoder, DBRNN = Deep Bidirectional Recurrent Neural Network, GANN = Graph Attention Network , SGD = Stochastic Gradient Descent, VAECF = Variational Autoencoder Collaborative Filtering, LR = Logistic Regression, SVM = Support Vector Machine, RF = Random Forest, DT = Decision Tree, NB = Naive Bayes, MLP = Multi-Layer Perceptron.} } 
  \label{tab:emotion}\\
  \toprule
  \textbf{Types} & \textbf{Papers} & \textbf{Data Source} & \textbf{Application} & \textbf{Techniques} & \textbf{Affective Terms} \\
  \midrule
  \endfirsthead
  \toprule
  \textbf{Types} & \textbf{Papers} & \textbf{Data Source} & \textbf{Application} & \textbf{Techniques} & \textbf{Affective Terms} \\
  \midrule
  \endhead
  \cmidrule(lr){2-6}
  \multicolumn{6}{r}{\textit{Continued on next page}} \\
  \midrule
  \endfoot

  \bottomrule
  \endlastfoot
    & \cite{deng2015musicemotion} & Music title, lyrics, time & Music & BPR, Cos & 21 emotion terms\\
    \cmidrule(lr){2-6}
    & \cite{moscati2024musicemotion} & Interactions, tags & Music & FM, DeepFM & Tenderness, joyful, nostalgia, wonder, sadness, tension \\
    \cmidrule(lr){2-6}
    & \cite{shen2020peia} & Interactions, tweets & Music & FM, DNN, DeepFM & More than 20 emotion terms \\
    \cmidrule(lr){2-6}
    & \cite{meng2018exploitemotion} & Ra, user ID & E-commerce & MF & Pos., Neg.   \\
    \cmidrule(lr){2-6}
    & \cite{BREITFUSS2021715} & Rev, item metadata & Movie & GraphDB & Apathy, joy, Neu., and Pos.   \\
    \cmidrule(lr){2-6}
    & \cite{ARAMANDA2023120190} & Rev, Ra & E-commerce & CF & 8 emotion terms   \\
    \cmidrule(lr){2-6}
    & \cite{poirson2019customeremotion} & Ra, Rev, item metadata & Movie & Pearson correlation & 8 emotion terms   \\  
    \cmidrule(lr){2-6}
    & \cite{LIM2017179} & Ra, Rev, item Des & Movie & Rule-based & Implicit  \\
    \cmidrule(lr){2-6}
    & \cite{zhang2024empathetic} & Rev, Ra & Movie & GPT, R-GCN, Llama-2 & 9 emotion terms   \\
    \cmidrule(lr){2-6}
    Cat. & \cite{kuo2005musicemotion} & Music metadata & Music & MAF, RW &  More than 20 emotion terms  \\
    \cmidrule(lr){2-6}
    & \cite{10.1145/3151759.3151817} & Tweets & Social media & SVM & 8 emotion terms   \\  
    \cmidrule(lr){2-6}
    & \cite{10.1145/2964284.2964327} & Tweets & Social media & SVM, LR, K-Means & Happiness, surprise, anger, disgust, fear, sadness   \\
    \cmidrule(lr){2-6}
    & \cite{10.1145/2072298.2072043} & Video, age, gender & Video & Rule-based & Fear, Neu., sadness, surprise, happiness, disgust, anger  \\
    \cmidrule(lr){2-6}
    & \cite{10.1145/3426020.3426119} & Music metadata, gender, age, interactions & Music, image & SVM, GA & Happy, sad, angry, surprised, bored, Neu.   \\
    \cmidrule(lr){2-6}
    & \cite{10.1145/3573834.3574478} & News metadata, interactions & News & CNN, Attn & Happy, angry, sad, surprise, fear   \\
    \cmidrule(lr){2-6}
    & \cite{10.1145/2700171.2791042} & Comments, user-item metadata & Short film & Cos, TF-IDF, PC & 8 emotion terms   \\  
    \cmidrule(lr){2-6}
    & \cite{Santos02012016} & physiological sensor data  & Education & Rule-based & Implicit  \\
    \cmidrule(lr){2-6}
    & \cite{guo2019collaborative} & Rev, Ra, item metadata & Movie & Rule-based & 14 emotion terms   \\
    \cmidrule(lr){2-6}
    & \cite{su2020emotioncloth} & Ra, video, gender, item images, item metadata & Fashion & RCNN, SVM & Happy, sad, angry, disgust, fear, surprise, Neu.  \\
    \cmidrule(lr){2-6}
    & \cite{lopez2021edurecomsys} & Ra, user-item metadata & Education & CF & 7 emotion terms   \\ 
    \cmidrule(lr){2-6}
    & \cite{adru2024musicemotion} & Images, music metadata & Music & LGBMClassifier, K-Means & Anger, fear, happy, disgust   \\
    \cmidrule(lr){2-6}
    & \cite{s21061997} & Audio, images & Music, image & SVM, GA & Happy, sad, angry, surprise, bored   \\  
    \cmidrule(lr){2-6}
    & \cite{7930506} & Video features & Video & SVM & 8 emotion terms   \\
    \cmidrule(lr){2-6}
    & \cite{7902195} & User-metadata, tweets & Social media & SGD, K-means & 6 emotion terms   \\
    
    \midrule
    & \cite{zhang2024affectivevideo} & Clicks, browsing, Ra, time & Short video & FM, DeepFM, AFM  & Happy, like, fear, sad, angry, hate, surprised, jealousy\\
    \cmidrule(lr){2-6}
    & \cite{KIM2018214} & Item Des, Ra & Education & MLE & Implicit   \\
    \cmidrule(lr){2-6}
    & \cite{costa2013emotion} & News text, user metadata & News & SVM, Rule-based & Implicit   \\
    \cmidrule(lr){2-6}
    & \cite{tkalvcivc2010using} & Ra, image metadata & Images & AdaBoost, NB, SVM & Implicit   \\  
    \cmidrule(lr){2-6}
    & \cite{10.3389/fpsyg.2019.00675} & User feedback & E-commerce & Rule-based & Implicit   \\
    \cmidrule(lr){2-6}
    & \cite{mizgajski2019affective} & Rev, news metadata & News & CF & 8 emotion terms   \\
    \cmidrule(lr){2-6}
    & \cite{CHHEDA2023383} & Music metadata, audio, image & Music & MobileNetV3, ResNet, EfficientNetB4 & Implicit   \\
    \cmidrule(lr){2-6}
    & \cite{https://doi.org/10.1155/2022/9548486} & Songs' tags & Music & MF & Implicit   \\
    \cmidrule(lr){2-6}
    Dim. & \cite{6378304} & Ra, images & Image & AdaBoost, KNN, NB, SVM & Joy, fear, anger, surprise, disgust, sadness   \\
    \cmidrule(lr){2-6}
    & \cite{6901531} & Ra, item metadata & Education & AdaBoost, NB, RF & 8 emotion terms   \\
    \cmidrule(lr){2-6}
    & \cite{ayata2018musicemotion} & GSR \& PPG signals & Music & DT, SVM, RF, K-NN & Implicit   \\
    \cmidrule(lr){2-6}
    & \cite{gonzalez2019deepemotion} & ECG \& GSR signals & Tourism & DCNN, SVM, K-NN & 7 emotion terms   \\  
    \cmidrule(lr){2-6}
    & \cite{10720053} & User metadata, news text & News & DT, NMF & Pos., Neg., Neu.   \\
    \cmidrule(lr){2-6}
    & \cite{tkalcic2013affectiveimages} & Video recording of users’ facial expressions, images & Images & K-NN, SVM, NB, AdaBoost & Joy, fear, anger, surprise, disgust, sadness   \\  
    \cmidrule(lr){2-6}
    & \cite{yoon2012musicemotion} & Ra, past interaction & Music & Rule-based & Angry, happy, sad, peaceful   \\
    \cmidrule(lr){2-6}
    & \cite{han2024musicemotion} & HRV \& GSR signals & Music & VAE, CNN, Attn & 8 emotion terms  \\
    \cmidrule(lr){2-6}
    & \cite{ferrato2022meta4rs} & Facial video & Tourism & Rule-based & Implicit   \\  
    \cmidrule(lr){2-6}
    & \cite{revathy2023lyemobert} & Music lyrics, audio clip & Music & BERT, NB, LR, SVM, RF & Happy, angry, sad, relaxed   \\
    \cmidrule(lr){2-6}
    & \cite{10.1145/2503385.2503484} & Image, acoustic features & Music & PCA & Implicit  \\  
    \midrule
    & \cite{deng2015emotional} & Ra, music metadata & Music & SVR, BR & 23 emotion terms\\
    \cmidrule(lr){2-6}
    & \cite{https://doi.org/10.1111/bjet.13209} & EEG \& GSR signals & Education & ResnetV2, CNN & Happy, Neu., sad, disgust, surprised, afraid, angry    \\
    \cmidrule(lr){2-6}
    \vtop{\hbox{\strut Dim. +}\hbox{\strut Cat.}\vspace{-1.2em}}& \cite{Zheng2016} & Ra, time, location, weather, user metadata & Movie & MF & Sad, Neu., happy, scared, surprise, angry, disgusted   \\  
    \cmidrule(lr){2-6}
    & \cite{ishanka2017prefiltering} & Rev, Ra, item metadata & Tourism & CF & 8 emotion terms   \\
    \cmidrule(lr){2-6}
    & \cite{tripathi2019emoware} & Video recording of users’ facial expressions, interactions & Video & SARSA, Q-Learning, DBRNN & Joy, sadness, fear    \\  
    \cmidrule(lr){2-6}
    & \cite{kim2024indoor} & Temperature, video & Smart home & GANN, SGD, MLP, MF & Implicit   \\
    \midrule
    \multirow{2}{*}{Latent} & \cite{rostami2024food} & Interactions, Ra, comments, ingredients & Food & MF, NeuMF, VAECF & Implicit   \\
    \cmidrule(lr){2-6}
    & \cite{yin2024collaborative} & Ra, Rev, past interaction & Movie & CF, MF & 14 emotion terms   \\  
\end{longtable}
}
\normalsize

\section{Mood-Aware Recommender Systems}
\label{sec:mood-rs}

Moods are prolonged affective states that influence user preferences and decision-making over time \cite{scherer2005emotions}. Unlike emotions, which are short-lived and linked to specific stimuli, moods persist for longer durations and can exert a more sustained influence on content consumption behaviors. Mood-aware recommender systems aim to enhance personalization by aligning recommendations with a user’s ongoing affective context extracted from signals such as behavioral cues (e.g., changes in browsing duration), physiological indicators (e.g., sustained elevation in skin conductance), contextual information (e.g., extended periods of rainy weather). Below we list representative mood-aware RS approaches from three major recommendation domains: music, food, and education.
\begin{itemize}
    \item {\bf Mood Modeling in Music}. \citeauthor{bontempelli2022flowmoods} \cite{bontempelli2022flowmoods} introduced Flow Moods, a mood-aware music recommendation system by Deezer that personalizes playlists based on six user-selected mood categories. Mood prediction, derived from curator-labeled data and audio embeddings, guided mood-filtered recommendations alongside collaborative filtering. Similarly, \citeauthor{chen2015mood} \cite{chen2015mood} proposed MoMusic, a hybrid system that infers mood from tempo and situation from lyrics using expert labels. Songs are matched to user preferences, collected through structured questionnaires, via rule-based filtering on features such as tempo and vocal style. \citeauthor{andjelkovic2016moodplay} \cite{andjelkovic2016moodplay} proposed MoodPlay, an interactive recommender that uses mood-aware filtering, audio similarity, and 2D mood-space visualization. Artist mood vectors are derived from Rovi metadata and organized using the Geneva Emotional Music Scale (GEMS) model. Users are represented by affective centroids, and recommendations are generated through profile- or trail-based exploration. In a follow-up study, \citeauthor{andjelkovic2019moodplay} \cite{andjelkovic2019moodplay} extended the system by introducing four interface variants with varying levels of affective control and experimentally showed that moderate interaction complexity best balances cognitive load, user satisfaction, and perceived recommendation quality. 

    \item {\bf Mood Modeling in Education}. In education, \citeauthor{tang2021improve} \cite{tang2021improve} developed a mood-adaptive recommender for online learners. Users are represented by feature vectors (e.g., degree, skills, preferences), and mood, categorized as positive, stable, or negative, guides recommendation strategy, such as suggesting challenging material for positive moods and simpler content for negative ones. However, the mood detection mechanism is unspecified.

    \item {\bf Mood Modeling in Recipe Recommendation}. \citeauthor{ueda2016recipe} \cite{ueda2016recipe} proposed a recipe recommendation system that models user mood along six dimensions, body, mental, taste, time, price, and modification, based on a lexicon of 1,758 mood-related words. Recipes are annotated on a [-5, +5] scale for each dimension, and users specify their current mood using sliders. Recommendations are generated by ranking recipes based on similarity to the user’s mood profile, with missing annotations inferred via cosine similarity from labeled recipes.
\end{itemize}
Table~\ref{tab:mood} lists representative mood-aware recommender system studies, organized according to their application domains, mood modeling strategies, and data sources. \\

{\scriptsize
\begin{longtable}{p{0.65cm}p{3.8cm}p{1.6cm}p{2cm}p{4cm}}
  \caption{{\bf Mood-aware} recommender systems papers. {\footnotesize\textit{Note:} EM = Expectation Maximization, Cos = Cosine Similarity, RF = Random Forest, CF = Collaborative Filtering, KNN = K Nearest Neighbor.}}
  \label{tab:mood}\\
  \toprule
  \textbf{Papers} & \textbf{Data Source} & \textbf{Application} & \textbf{Techniques} & \textbf{Affective Terms} \\
  \midrule
  \endfirsthead
  \toprule
  \textbf{Papers} & \textbf{Data Source} & \textbf{Application} & \textbf{Techniques} & \textbf{Affective Terms} \\
  \midrule
  \endhead
  \multicolumn{5}{r}{\textit{Continued on next page}} \\
  \midrule
  \endfoot
  \bottomrule
  \endlastfoot
    \cite{bontempelli2022flowmoods} & User explicit preference, audio & Music & VGG-like, RF, CF & 7 mood terms \\
    \midrule
    \cite{chen2015mood} & Song metadata & Music & CF & Implicit\\
    \midrule
    \cite{andjelkovic2016moodplay} & Artist metadata, audio metadata & Music & K-NN & Implicit \\
    \midrule
    \cite{andjelkovic2019moodplay} & Artist metadata, audio metadata & Music & RF, VGG-like & 7 mood terms \\
    \midrule
    \cite{marshall2016mood} & Tweets & Social media & EM & Implicit \\
    \midrule
    \cite{ueda2016recipe} & Recipe, rating & Food & EM, CF & Cheerful, exhilarated\\
    \midrule
    \cite{tang2021improve} & Past interaction, user metadata & Education & Cos & Pos, Neg, stable \\
\end{longtable}
}
\normalsize



\section{Hybrid Affective Recommender Systems}
\label{sec:hybrid-rs}

\begin{figure}[t]
\centering
\begin{tikzpicture}[
  box/.style={circle, rectangle, draw, rounded corners, minimum width=0.05cm, minimum height=0.4cm,  minimum size=0.05cm, align=center, font=\footnotesize, fill=#1!20},
  line/.style={draw, thick},
]

\node[box=blue] (hybrid) at (0,0) {Hybrid};

\node[box=blue] (att_emo)     at ($(hybrid)+(2.5,0.4)$) {Attitude + Emotion};
\node[box=blue] (emo_mood)     at ($(hybrid)+(2.5,-0.4)$) {Emotion + Mood};

\node[box=blue] (serndipity)   at ($(att_emo)+(3,0)$) {Attitude + Surprise\\$\hookrightarrow$ Serendipity};

\node[box=blue] (dis)     at ($(serndipity)+(3,0.35)$) {Discovery};
\node[box=blue] (con)   at ($(serndipity)+(3,-0.35)$) {Content};

\draw[line] (hybrid.east) -- ++(0.5,0) |- (att_emo.west);
\draw[line] (hybrid.east) -- ++(0.5,0) |- (emo_mood.west);

\draw[line] (att_emo.east) -- ++(0.5,0) |- (serndipity.west);

\draw[line] (serndipity.east) -- ++(0.5,0) |- (dis.west);
\draw[line] (serndipity.east) -- ++(0.5,0) |- (con.west);

\end{tikzpicture}
\caption{Taxonomy for hybrid affective recommender systems. Representative papers for each of the 2 main hybrid categories are listed in Table~\ref{tab:hybrid}.}
\label{fig:hybrid-taxonomy}
\end{figure}
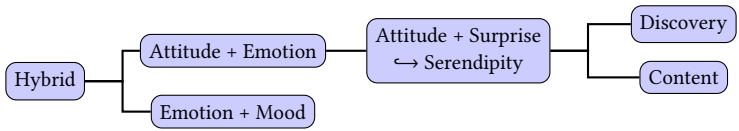

Hybrid affective recommender systems integrate multiple types of affective states, such as sentiment, emotion, and mood, to leverage their strengths and mitigate the limitations of using a single type alone. By combining these different affective dimensions, such systems aim to enhance personalization, improve predictive accuracy, and increase overall user satisfaction. We categorize hybrid approaches in two main categories based on their methodological integration of affective state types: {\bf attitude + emotion}, which combine long-term evaluative judgments with short-term affective reactions, and {\bf emotion + mood}, which integrate emotional responses with mood to capture both momentary and enduring aspects of user preferences. Figure~\ref{fig:hybrid-taxonomy} represents the taxonomy for hybrid affective recommender systems. Table~\ref{tab:hybrid} provides a structured list of representative papers in the two main categories, highlighting their targeted affective states, fusion strategies, application domains, and data sources.

{\scriptsize
\begin{longtable}{p{0.9cm}p{0.65cm}p{3.5cm}p{1.3cm}p{2.35cm}p{3.0cm}} 
  \caption{{\bf Hybrid} recommender systems papers. {\footnotesize\textit{Note:} Rev = Reviews, Rat = Ratings, MF = Matrix Factorization, NMF = Non-Negative Matrix Factorization, KL div = Kullback–Leibler Divergence, SGD = Stochastic Gradient Descent, LDA = Latent Dirichlet Allocation, CS = Cosine Similarity, LR = Logistic Regression, MLP = Multi-Layer Perceptron, CNN = Convolutional Neural Network, KNN = K Nearest Neighbor,  FM = Factorization Machine, DeepFM	= Deep Factorization Machine, AFM = Attentional Factorization Machine, SVD = Singular Value Decomposition, PCA = Principal Component Analysis, CF = Collaborative Filtering.}} 
  \label{tab:hybrid}\\
  \toprule
  \textbf{Types} & \textbf{Papers} & \textbf{Data Source} & \textbf{Application} & \textbf{Techniques} & \textbf{Affect Modeling / Terms} \\
  \midrule
  \endfirsthead
  \toprule
  \textbf{Types} & \textbf{Papers} & \textbf{Data Source} & \textbf{Application} & \textbf{Techniques} & \textbf{Affective Representation} \\
  \midrule
  \endhead
  \multicolumn{6}{r}{\textit{Continued on next page}} \\
  \midrule
  \endfoot

  \bottomrule
  \endlastfoot
    & \cite{wang2023airline} & Rev & Airline & EmoLex & Discrete / 8 emotion terms \\
    \cmidrule(lr){2-6}
    & \cite{sertkan2022emotion} & Click, interactions, news text & News & BERT & Discrete / 28 emotion terms \\
    \cmidrule(lr){2-6}
    \makecell[lt]{Emotion+ \\\hspace*{\fill}Attitude} & \cite{chen2018emotional} & Lyric & Music & CF, TF-IDF & Implicit / happy, angry, sad, relaxing \\
    \cmidrule(lr){2-6}
    & \cite{10.1145/3627043.3659561} & Interactions, music metadata & Music & FM, DeepFM, NFM & Implicit \\
    \cmidrule(lr){2-6}
    & \cite{https://doi.org/10.1155/2022/7246802} & Rev, Ra, user-item metadata & E-commerce & SVM, Apriori algorithm & Implicit / happy, sad, relax, fear \\
    \midrule
    
    & \cite{gilda2017musicmoodemotion} & Images, item metadata & Music & CNN, ANN, SGD, Cos & Happy, sad, angry,  \\
    \cmidrule(lr){2-6}
    & \cite{cai2007emotion} & Lyric, review & Music & LDA, KL div & 40 emotion terms \\
    \cmidrule(lr){2-6}
    \makecell[lt]{Emotion\\ \hspace*{\fill}+ Mood} & \cite{Zheng2016} & Ra, time, location, weather, user metadata & Movie & MF & Sad, happy, scared, surprise, angry, disgusted, neutral \\
    \cmidrule(lr){2-6}
    & \cite{piazza2017affective} & Ra, age, gender, item image & Fashion & FM, SGD & 23 emotion terms \\
    \cmidrule(lr){2-6}
    & \cite{polignano2021emotion} & Rev, audio metadata, lyric, time & Music & Cos, LR & Joy, anger, sadness, surprise \\
    \cmidrule(lr){2-6}
    & \cite{LIU2023103256} & Audio metadata, genre, interactions & Music & LSTM, CNN, KNN & Happy, anger, sad, fear \\
\end{longtable}
}
\normalsize

\noindent 


\subsection{Attitude + Emotion Informed Recommender Systems} 

Several studies combine attitude and emotion to model human affective states in recommender systems in order to capture nuanced affective interactions. For instance, \citeauthor{wang2023airline} \cite{wang2023airline} analyzed airline reviews in order to assess the impact of sentiment and emotion on recommendation intention. Positive sentiment, joy, and trust were found to increased the likelihood of recommendation, while negative sentiment, anger, and disgust had the opposite effect. \citeauthor{sertkan2022emotion} \cite{sertkan2022emotion} proposed a news recommender system that integrates semantic and emotional signals from both articles and user behavior, based on sentiment and Ekman's emotion taxonomy. \citeauthor{chen2018emotional} \cite{chen2018emotional} introduced a music recommender that extracts sentiment polarity and arousal-valence (AV) features from lyrics using a Chinese sentiment lexicon. Songs are represented in affective space using emotion point matrices and matched to mood-specific AV regions for recommendation.

A notable subclass of emotion-attitude hybrid recommender systems targets {\bf serendipity}, a well-established concept in RS literature that is not always explicitly framed in affective terms. Serendipity is commonly described as an {\it unexpected} yet {\it pleasant} experience. This characterization reveals its underlying affective structure: the unexpectedness component corresponds to the {\it emotion} of surprise, while pleasantness reflects a positive evaluative {\it attitude} toward the outcome. From this perspective, serendipity can be understood as a hybrid affective experience arising from the interplay between emotional arousal and attitudinal appraisal. 

\subsubsection{\bf Cognitive and Affective Foundations of Serendipity} Modern perspectives emphasize that serendipity arises from a confluence of cognitive and emotional processes. \citeauthor{andel1994serendipity} \cite{andel1994serendipity} argued that unsought findings occur when a well-prepared mind encounters an unforeseen stimulus and, through careful interpretation, transforms it into something valuable. \citeauthor{marg1995descartes} \cite{marg1995descartes} demonstrated how emotional responses serve as somatic markers that guide decision-making and flag surprising events. In a similar vein, \citeauthor{isen2001affect} \cite{isen2001affect} provided empirical evidence that positive affect not only enhances cognitive flexibility and fosters creative associations but also increases our sensitivity to unexpected stimuli. \citeauthor{foster2003serendipity} \cite{foster2003serendipity} further supported the cognitive-affective perspective on serendipity in information seeking context by demonstrating that exploratory search behaviors can yield useful, unexpected information. More recently, \citeauthor{busch2024serendipity} \cite{busch2024serendipity} highlighted that conditions such as agency, surprise, and value are essential for serendipitous outcomes, suggesting that individuals with rich prior experience and an emotionally engaged mindset are better equipped to encounter and leverage the unexpected. Serendipity emerges from the dynamic interplay between a prepared, flexible mind and the affective responses triggered by unexpected stimuli, with positive emotions facilitating creative insight and novel discoveries. 

\subsubsection{\bf Serendipity in Information Retrieval and RS} The concept of serendipity in recommender systems originated from early work in information retrieval, where researchers observed that users occasionally encountered unexpectedly useful information while navigating large, unstructured datasets \cite{Toms2000SerendipitousIR}. Such studies highlighted the paradox of designing systems, because aiming to produce unexpected discoveries makes them expected, that deliberately facilitate unexpected discoveries rather than merely optimizing for expected information needs \cite{Toms2000SerendipitousIR, foster2003serendipity, mcbirnie2008seeking}. 

\citeauthor{herlocker2004collaborative} \cite{herlocker2004collaborative} informally define a serendipitous recommendation as one that helps the user find a "surprisingly interesting item" that they might not have otherwise discovered, emphasizing surprise in delivering unexpected items. Building on this, subsequent studies have refined the definition in various ways. For instance, \citeauthor{iaquinta2008serendipity} \cite{iaquinta2008serendipity} proposed that an item might be considered serendipitous if a classifier is uncertain about its relevance, while \citeauthor{adamopoulos2014unexpectedness} \cite{adamopoulos2014unexpectedness} argued that an item is serendipitous if it markedly deviates from the user’s established profile. Yet, despite these advances, there is still no consensus on the definition of serendipity in recommender systems. While many agree that serendipity should incorporate surprise, novelty, and utility, the relative importance of these components varies. Some researchers argue that for a recommendation to be serendipitous, it must be entirely novel, implying that the user is completely unaware of the item, whereas others contend that it is sufficient for the recommendation to be unexpected relative to the user’s current preferences, even if the user might eventually have discovered it independently \cite{kaminskas2014measuring}.

Two distinct forms of serendipity have been notably observed in the literature, based on the source from which the serendipitous experience emerges:
\begin{itemize}
    \item In {\bf discovery-level} serendipity, users stumble upon entirely unfamiliar items, such as new genres, categories, or creators.
    \item In contrast, {\bf content-level} serendipity arises not from how the item was found, but from unexpected features or characteristics encountered during its consumption. These features may relate to topics, style, or any other aspect of the item whose engaging or valuable nature becomes apparent only through direct experience. 
\end{itemize}
Table~\ref{tab:serendipity} lists representative papers on serendipity-aware recommender systems, highlighting their data sources, application domains, and techniques.

\subsubsection{\bf Discovery-Level Serendipity in RS} Discovery-level serendipity refers to user experiences where the recommended item is not have been actively sought by the user. This form of serendipity emphasizes the \textit{stumbling-upon effect}, where users are exposed to items that lie outside their established preference boundaries yet turn out to be rewarding or useful. This conceptualization of serendipity is reflected in early foundational work on recommender systems. For example, \citeauthor{herlocker2004collaborative} \cite{herlocker2004collaborative} considers that ``\textit{A serendipitous recommendation helps the user find a surprisingly interesting item he might not have otherwise discovered.}'' This framing highlights the role of unintentional discovery and positions serendipity as orthogonal to pure accuracy, underscoring the value of systems that promote exploration beyond familiar content. A similar conceptualization can be observed in the question below taken from a post-study questionnaire administered by \citeauthor{taramigkou2013escape} \cite{taramigkou2013escape} for their guided music-exploration system:

\begin{center}
\begin{tcolorbox}[width=0.95\textwidth,
    colback=white,
    colframe=gray!10!black, 
    colbacktitle=black!60, 
    coltitle=white,        
    title=Post-study Question,
    fonttitle=\bfseries,
    boxrule=0.5pt,
    left=4pt,
    right=4pt,
    top=2pt,
    bottom=2pt,
    enhanced,
    sharp corners,
    before skip=6pt,
    after skip=6pt]
\small
``\textit{Did you find artists you wouldn’t have found easily on your own and which you would like to listen to from now on?}"
\end{tcolorbox}
\end{center}
In more recent work, \citeauthor{fu2023serendipity} \cite{fu2023serendipity} constructed a ground-truth serendipity dataset by identifying reviews in which users explicitly mention stumbling upon an item. These reviews were collected through a crowd-sourced annotation process and reflect real-world accounts of unexpectedly discovering and enjoying an item. One example review from their dataset is shown below:

\begin{center}
\begin{tcolorbox}[width=0.95\textwidth,
    colback=white,
    colframe=gray!10!black, 
    colbacktitle=black!60, 
    coltitle=white,        
    title=Example Review,
    fonttitle=\bfseries,
    boxrule=0.5pt,
    left=4pt,
    right=4pt,
    top=2pt,
    bottom=2pt,
    enhanced,
    sharp corners,
    before skip=6pt,
    after skip=6pt]
\small
``\textit{I {\bf stumbled on} this book by accident, it was on an iPad I had borrowed and was checking out the features... but it grabbed my attention from the first page and I could not put it down. I recommended it to friends who loved it just as much. It is not my usual genre, but I loved it. It had everything—tragedy, suspense, romance, and an interesting backdrop for a story.}''
\end{tcolorbox}
\end{center}
This review illustrates the defining characteristics of discovery-level serendipity: the item lies outside the user’s routine preferences, the discovery is incidental, and the emotional outcome is strongly positive. Building on this dataset, subsequent work explored serendipity in a cross-domain recommendation setting, introducing a deep learning model designed to facilitate unexpected discoveries across the book and movie domains \cite{fu2024serendipity}. In addition, several studies have explored discovery-level serendipity using only coarse signals like item IDs, user-item interactions, and popularity, without analyzing textual, visual, or other rich content features \cite{degemmis2015serendipity, li2020unexpectedness, li2020directional, lu2012serendipitous, afridi2018serendipity, adamopoulos2014unexpectedness}.

Collectively, these studies demonstrate that discovery-level serendipity, which stems from exposure to unfamiliar yet rewarding content, can be achieved through recommendation strategies that prioritize item or category level unfamiliarity.

\subsubsection{\bf Content-Level Serendipity in RS} Content-level serendipity refers to cases where users are recommended items that have internal features that are both unexpected and useful. Unlike discovery-level serendipity, which emphasizes stumbling upon unfamiliar items, content-level serendipity emerges from engaging with an item more deeply and uncovering surprising aspects that are not obvious before consuming the item, as illustrated in the user review below:

\begin{center}
\begin{tcolorbox}[width=0.95\textwidth,
    colback=white,
    colframe=gray!10!black, 
    colbacktitle=black!60, 
    coltitle=white,        
    title=Example Review,
    fonttitle=\bfseries,
    boxrule=0.5pt,
    left=4pt,
    right=4pt,
    top=2pt,
    bottom=2pt,
    enhanced,
    sharp corners,
    before skip=6pt,
    after skip=6pt]
\small
``\textit{It seemed like just a basic travel camera, something light to carry around. But the moment I used the touch-to-focus and silent shutter mode, which {\bf blew me away}, it felt like I was holding something far more premium. {\bf I didn’t expect} something this compact to feel so refined and professional. I’m loving it more and more each day.}''
\end{tcolorbox}
\end{center}
In this example, the user’s initial expectations were determined by the product’s lightweight design and intended use case. These expectations were confounded upon using the item, when serendipity arose from unexpectedly discovering high-end features through direct interaction. This highlights how content-level serendipity is driven by internal characteristics that positively surprise the user during or after item consumption.

\citeauthor{hasan2023serendipity} \cite{hasan2023serendipity} introduced a formal definition that equates content-based serendipity with the product between a user's rating of an item and Bayesian surprise \cite{itti_bayesian_2005} quantified as the Kullback-Leibler (KL) divergence between the prior (before consuming the item) and the posterior (after consuming the item) distributions over the user’s topic-level preferences. To recommend items with high potential for serendipity, the systems first uses a collaborative filtering approach to identify users who, at some prior point in their reading history, had similar topic-level preferences to the target user. Items with high serendipity scores from the most similar users are then recommended to the target user. Other RS approaches to content-level serendipity, such as \cite{niu2018surprise,jenders2015serendipity}, quantify surprise more directly as a semantic dissimilarity between the recommended item and a user's historical preference profile. When topic modeling is used to model content, dissimilarity can be quantified as the KL divergence between the candidate item and the user's historical topic distributions \cite{huang2018learning}. Apart from these works, the user study of \citeauthor{kotkov2018serendipity} \cite{kotkov2018serendipity} explicitly evaluates both content‑level and discovery-level serendipity, asking participants separate questions about whether the recommended item was something they would not normally discover and whether its style, genre, or topic differed markedly from their typical choices.

Together, these approaches underscore that content-level serendipity arises not merely from unfamiliarity, but from meaningful deviations within an item's content or features that defy user expectations in rewarding ways, which highlights the importance of modeling rich item information. \\

{\scriptsize
\begin{longtable}{p{1cm}p{0.9cm}p{4.2cm}p{2.2cm}p{3.4cm}} 
  \caption{{\bf Serendipity-aware} recommender systems papers. {\footnotesize\textit{Note:} Rev = Reviews, Ra = Ratings, MF = Matrix Factorization, NMF = Non-Negative Matrix Factorization, GMM = Gaussian Mixture Model, CapsNets = Capsule Networks, BisoNet =  Bisociative Information Network, MC = Markov Chain, BLR = Bayesian Linear Regression, AROW = Adaptive Regularization of Weight, VAE = Variational Autoencoder, NMF = Non-negative Matrix Factorization, GMM = Gaussian Mixture Model, RW = Random Walk, KL div = Kullback–Leibler Divergence, SGD = Stochastic Gradient Descent, Pos = Positive, Neu = Neutral, Neg = Negative, LDA = Latent Dirichlet Allocation, CS = Cosine Similarity, LR = Logistic Regression, MLP = Multi-Layer Perceptron, CNN = Convolutional Neural Network, KNN = K Nearest Neighbor,  FM = Factorization Machine, DeepFM	= Deep Factorization Machine, AFM = Attentional Factorization Machine, SVD = Singular Value Decomposition, PCA = Principal Component Analysis, CF = Collaborative Filtering, PMI = Pointwise Mutual Information.}} 
  \label{tab:serendipity}\\
  \toprule
  \textbf{Types} & \textbf{Papers} & \textbf{Data Source} & \textbf{Application} & \textbf{Techniques} \\
  \midrule
  \endfirsthead
  \toprule
  \textbf{Types} & \textbf{Papers} & \textbf{Data Source} & \textbf{Application} & \textbf{Techniques} \\
  \midrule
  \endhead
  \multicolumn{5}{r}{\textit{Continued on next page}} \\
  \midrule
  \endfoot

  \bottomrule
  \endlastfoot
    & \cite{yang2017improving} & Ra & Movie & MF  \\
    \cmidrule(lr){2-5}
    & \cite{ge2020serendipity} & Interactions, location & Tourism & RW, Word2Vec  \\
    \cmidrule(lr){2-5}
    & \cite{zhang2021next} & Interactions, location & Tourism & Transformer \\
    \cmidrule(lr){2-5}
    & \cite{ZIARANI2021115660} & Ra & Movie & CNN \\
    \cmidrule(lr){2-5}
    & \cite{boo2023session} & Click, browsing history & E-commerce & GNN \\
    \cmidrule(lr){2-5}
    & \cite{afridi2018serendipity} & Ra & Education & Jacquard similarity \\
    \cmidrule(lr){2-5}
    & \cite{fu2023serendipity} & Rev & Book & Transformer \\
    \cmidrule(lr){2-5}
    & \cite{xu2020serendipity} & Ra, interactions & Movie, book & MLP, MF,  \\
    \cmidrule(lr){2-5}
    & \cite{li2019serendipity} & Ra, movie metadata & Movie & RNN \\
    \cmidrule(lr){2-5}
    Discovery & \cite{adamopoulos2014unexpectedness} & Ra, item tag & Book & MF, KNN \\
    \cmidrule(lr){2-5}
    & \cite{fu2024serendipity} & Rev & Movie, book & MF, BERT \\
    \cmidrule(lr){2-5}
    & \cite{li2020unexpectedness} & Ra, click, interactions & Movie, restaurant & MLP, GRU \\
    \cmidrule(lr){2-5}
    & \cite{li2020unexpected} & Ra, Rev, interactions & Tourism & NCF, KNN, FM, AE \\
    \cmidrule(lr){2-5}
    & \cite{zhang2012serendipity} & Interactions & Music & LDA \\
    \cmidrule(lr){2-5}
    & \cite{onuma2009surprise} & Ra & Movie & RW \\
    \cmidrule(lr){2-5}
    & \cite{chen2019serendipity} & User survey data & E-commerce & CF \\
    \cmidrule(lr){2-5}
    & \cite{zheng2015unexpectedness} & Ra & Movie & Rule-based \\
    \cmidrule(lr){2-5}
    & \cite{wang2020impacts} & Clicks, interactions, item metadata & E-commerce & Apriori algorithm, LR \\
    \cmidrule(lr){2-5}
    & \cite{pandey2018recommending} & Ra & Movie & NCF, NeuMF, MLP \\
    \cmidrule(lr){2-5}
    & \cite{li2020directional} & Ra, interactions & Book, movie & GMM, CapsNets \\
    \cmidrule(lr){2-5}
    & \cite{lee2020serendipity} & Interactions, app description & Mobile app & VAE \\
    \cmidrule(lr){2-5}
    & \cite{taramigkou2013escape} & Interactions, music tag, artist & Music & LDA, Cos  \\
    \cmidrule(lr){2-5}
    & \cite{kawamae2010serendipitous} & Item metadata, Ra & Movie, music & MC  \\
    \cmidrule(lr){2-5}
    & \cite{lu2012serendipitous} & Ra & Music, movie & SVD  \\
    \cmidrule(lr){2-5}
    & \cite{murakami2007metrics} & Item metadata & TV program & Bayesian network  \\
    \cmidrule(lr){2-5}
    & \cite{wang2023item} & Ra, click, interactions, user-item metadata & E-commerce, Movie & PCA, Spearman’s correlation \\
    \cmidrule(lr){2-5}
    & \cite{8392508} & Interactions, browsing & E-commerce & KNN, Cos  \\
    \cmidrule(lr){2-5}
    & \cite{grange2019little} & Rev, Ra, interactions & Restaurant & LR  \\

    \midrule
    
    & \cite{sugiyama2015towards} & User publication history, citation & Research article & Clustering, CF  \\
    \cmidrule(lr){2-5}
    & \cite{afridi2020facilitating} & User metadata & Research article &   \\
    \cmidrule(lr){2-5}
    & \cite{niu2021luckyfind} & News topic, Ra & Health news & PMI, Rule-based  \\
    \cmidrule(lr){2-5}
    & \cite{maake2019serendipitous} & Paper title & Education & BisoNet, Log-likelihood Ratio \\
    \cmidrule(lr){2-5}
    & \cite{niu2018adaptive} & News topic & Health news & Probabilistic method \\
    \cmidrule(lr){2-5}
    & \cite{jenders2015serendipity} & News title, text, news metadata & News & LDA, Cos \\
    \cmidrule(lr){2-5}
    Content & \cite{hasan2023serendipity} & Ra, Rev, interactions & Book & BLR, AROW \\
    \cmidrule(lr){2-5}
    & \cite{li2024variety} & Ra, interactions & E-commerce, Movie & FM, NCF, KNN, NMF \\
    \cmidrule(lr){2-5}
    & \cite{li2020utility} & Interactions & E-commerce & Rule-based \\
    \cmidrule(lr){2-5}
    & \cite{fan2018implementing} & News topic, click & Health news & Rule-based \\
    \cmidrule(lr){2-5}
    & \cite{niu2018surprise} & Click, news topic  & Health news & LDA, KL div  \\
    \cmidrule(lr){2-5}
    & \cite{huang2018learning} & Click, item description & Entity & CNN, LDA \\
    \cmidrule(lr){2-5}
    & \cite{kotkov2020does} & Ra, serendipity label & Movie & SVD  \\
    \cmidrule(lr){2-5}
    & \cite{niu2017framework} & News topic, user feedback & Health news & Rule-based  \\
    \cmidrule(lr){2-5}
    & \cite{maccatrozzo2017sirup} & Item title, genre & TV program & Cos, LR  \\






\end{longtable}
}
\normalsize

\subsection{Emotion + Mood Informed Recommender Systems} 

Integrating information about transient states (emotions) with more sustained states (moods) allows recommender systems to adapt to both short-term affective contexts and more persistent preference patterns. An example in this category is the \textit{MusicSense} framework \cite{cai2007emotion}, which aims to match the emotions and moods expressed by the music with those identified in web pages that the user is reading. Based on a generative probabilistic model inspired by LDA, both music and textual content are represented as mixtures of emotions and moods. The similarity between music and web content is assessed using KL divergence, resulting in recommendations matching emotional experiences and mood tendencies. In a related study, \citeauthor{piazza2017affective} \cite{piazza2017affective} leveraged moods measured via the Positive and Negative Affect Schedule (PANAS) alongside emotions captured through the Pleasure-Arousal-Dominance (PAD) model within a factorization machines framework. Their findings indicated that mood features significantly enhanced predictive accuracy, especially in cold-start scenarios, whereas emotions were found to introduce noise due to their transient nature, diminishing their predictive utility for product evaluation. \citeauthor{moscato2021musicemotion} \cite{moscato2021musicemotion} developed an emotion-aware music recommender that leverages song-derived emotions to continuously update the user moods represented within a reduced PAD space, while item recommendation is performed by identifying the nearest neighbors to the user’s current mood. \citeauthor{dhahri2018moodaware} \cite{dhahri2018moodaware} proposed a mood-aware music recommender that infers user mood (positive, negative, or neutral) from social media cues and combines it with adaptive song embeddings in a 2D latent space to deliver personalized recommendations without requiring explicit input or listening history. Mood-specific song relevance is computed via cosine similarity between user and song emotion vectors, refined through reinforcement learning, and updated using change-point detection. 

\section{Datasets and Applications}

\label{sec:dataset_application}

Affective recommender systems leverage a wide variety of datasets and have been applied across multiple domains to enhance user experience. Unlike traditional RS, which primarily rely on explicit user feedback (e.g., ratings, clicks, purchase history), affective RS incorporate emotion, mood, and sentiment-related information from various data sources. These include physiological signals (e.g., EEG, ECG, GSR), text, multimodal interactions, and behavioral cues that help infer users’ affective states. This section explores the key datasets used in affective RS and provides an overview of their applications across different domains. 

\subsection{Datasets for Affective Recommender Systems}

Affective RS datasets originate from various sources, including social media interactions, movie reviews, music streaming histories, fashion preferences, and conversational dialogues. Some datasets are explicitly designed for affective recommendations, while others are adapted from general recommender system datasets with additional affective annotations. Table~\ref{tab:affective_datasets} provides an overview of key datasets for affective RS, covering multiple domains such as short video recommendations, social media, e-commerce, fashion, music, healthcare, and news.

These datasets illustrate the growing use of affective information across recommendation models in various domains. Nevertheless, the subjective nature of affective states introduces annotation challenges, often leading to inconsistent emotion labels \cite{8764449}. Additionally, most existing datasets are limited to single-modality signals, which is limiting for multimodal affective modeling. Consequently, ideas for future research include the development of large-scale multimodal datasets that integrate diverse sources such as textual reviews, contextual data, vision, audio, and physiological signals. Leveraging self-supervised learning and weakly supervised annotation techniques can further facilitate creating comprehensive affect-rich datasets. Tackling these data challenges can help affective RS to more accurately adapt to users' dynamic affective states, providing more personalized and ultimately more satisfying recommendations.

\subsection{Applications of Affective Recommender Systems} 

\begin{figure}[t]
    \centering
    \includegraphics[width=0.8\linewidth]{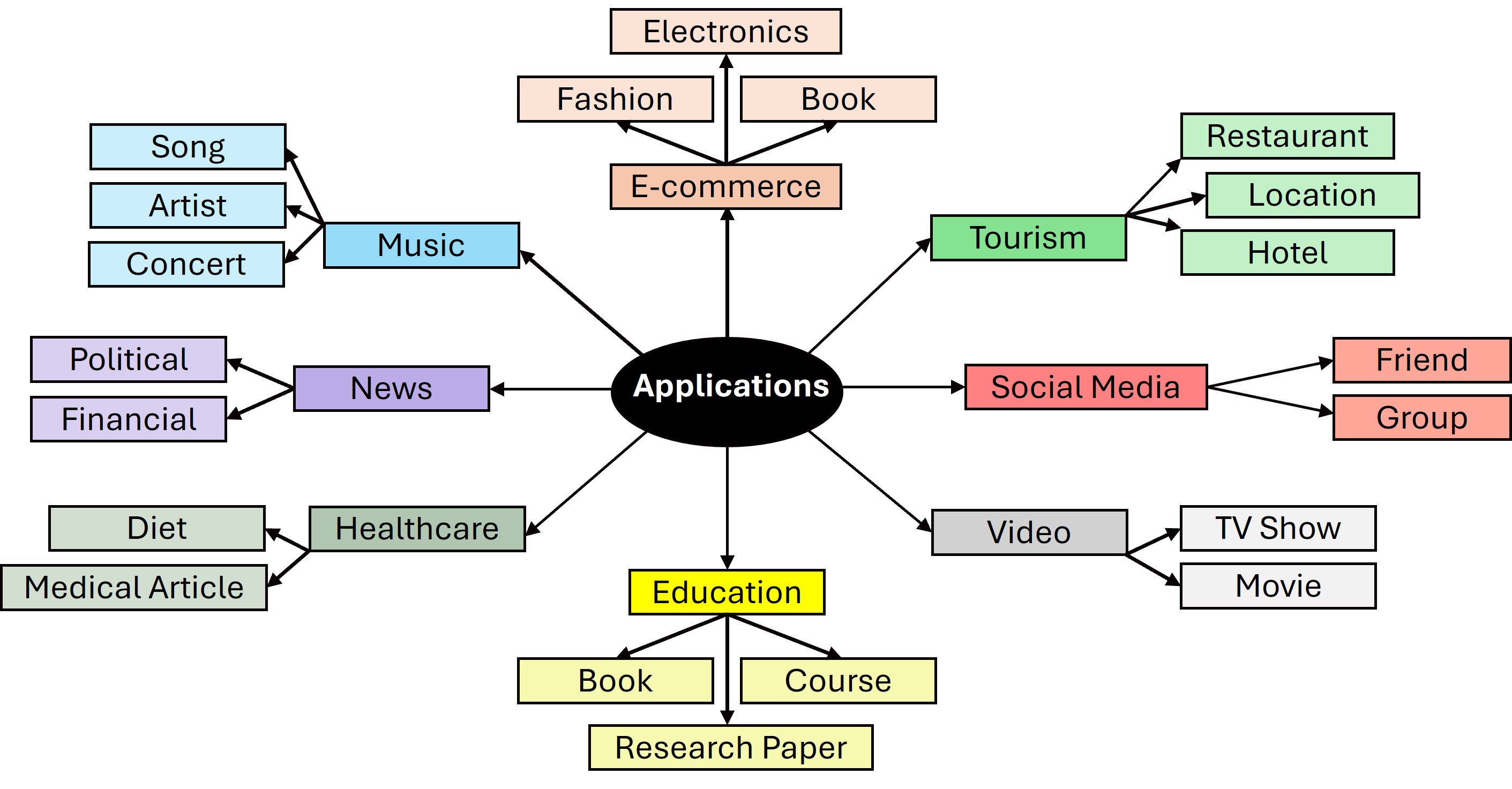}
    \caption{Applications of affective recommender systems.}
    \label{fig:application}
\end{figure}

Affective recommender systems are increasingly being adopted across many domains. Fig.~\ref{fig:application} presents the primary domains and their respective sub-domains where affective RS have been notably applied.
\begin{itemize}
    \item {\bf E-commerce}. In e-commerce, affective RS analyze user sentiment from product reviews, emotional feedback, and purchase behavior to personalize recommendations in areas such as electronics, fashion, and books \cite{ARAMANDA2023120190, 10.1145/3404835.3462943, cai2022deepsentiment, meng2018exploitemotion}.
    \item {\bf Education}. Educational platforms integrate affective signals from social media and micro-blogs to dynamically personalize learning materials, including course recommendations, research articles, and educational books, helping students stay engaged and motivated based on their emotional states \cite{n2023coursesentiment, Santos02012016, 10184251, https://doi.org/10.1111/bjet.13209}.
    \item {\bf Healthcare}. In the healthcare sector, affective RS play a vital role in tailoring health-related suggestions, such as personalized diet plans, hospital and drug recommendations, and useful medical articles. By recognizing patterns of emotional distress, these systems support improved health outcomes and emotional well-being \cite{gonzalez2019deepemotion, SERRANOGUERRERO2024122340, shi2022sengr}.
    \item {\bf Music}. Music streaming services employ emotion-aware algorithms to adjust playlists and recommend songs, artists, or concerts that resonate with a user’s current mood, enriching their listening experience \cite{10.1145/2503385.2503484, kuo2005musicemotion, deng2015emotional, revathy2023lyemobert, han2024musicemotion}.
    \item {\bf Tourism}. Tourism recommendation systems incorporate affective insights from user reviews and multimodal interactions to suggest personalized travel experiences, including restaurant choices, hotel accommodations, airline services, and travel destinations \cite{artemenko2020using, 10.1145/2481492.2481505, 10.1145/2442810.2442816, ferrato2022meta4rs, asani2021restaurantsentiment}.
    \item {\bf Video}. In video streaming services, affective RS adapt movie, TV show, and short video recommendations based on users’ emotional preferences and past viewing behavior, ensuring a more engaging and immersive content selection process \cite{zhang2024affectivevideo, 10.1145/2072298.2072043, 6851912}.
    \item {\bf Social Media}. Social media platforms utilize affective engagement patterns, such as sentiment in posts, reactions, and interactions, to recommend potential friends, communities, and interest-based groups \cite{10.1145/2964284.2964327, 10.1145/3151759.3151817, 10.1145/3534678.3539054}.
    \item {\bf News}. News recommender systems dynamically adapt content delivery based on users' emotional reactions to political, financial, and global events, ensuring engagement while mitigating content fatigue or emotional overload \cite{10.1145/3573834.3574478, 10720053, mizgajski2019affective}. 
\end{itemize}


{\footnotesize
\begin{table}[htbp]
\centering
\caption{Available datasets across different domains for affective recommender systems.}
\label{tab:affective_datasets}
\begin{tabular}{p{1.3cm}p{1.8cm}p{9.1cm}}
    \toprule
    \textbf{Domain} & \textbf{Dataset Name} & \textbf{Description} \\
    \midrule
    & \vtop{\hbox{\strut ReDial}\hbox{\strut Li et al. \cite{NEURIPS2018_800de15c}}} & The ReDial dataset contains over 10,000 two-party dialogues with 182,150 utterances covering 51,699 movies. It includes user feedback (“like,” “dislike,” “not say”) for precise evaluation and sentiment-labeled responses, enabling sentiment-aware recommendations.\\ 
    \cmidrule(lr){2-3}
    Movie & \parbox[t]{3.5cm}{\raggedright MovieLens 100K, \\1M, 10M,  20M\\ Harper et al. \cite{10.1145/2827872}} 
    & A suite of movie rating datasets containing between 100K and 20M ratings from 943 to 138,493 users on 1,682 to 27,278 movies. \\
    \cmidrule(lr){2-3}
    & \vtop{\hbox{\strut Netflix}\hbox{\strut Bennett et al. \cite{bennett2007netflix}}\vspace{-1.2em}} & The Netflix Prize dataset consists of over 100 million movie ratings collected from 480,000 anonymous users across 17,770 movie titles between October 1998 and December 2005. \\
    \midrule
    Fashion & \vtop{\hbox{\strut Amazon Fashion}\hbox{\strut He et al. \cite{10.1145/2872427.2883037}}\vspace{-1.2em}} & A large-scale dataset from Amazon consisting of two sub-datasets: Women’s Clothing \& Accessories and Men’s Clothing \& Accessories, spanning from March 2003 to July 2014.\\ 
    \cmidrule(lr){2-3}
    & Piazza et al. \cite{piazza2017affective} & A novel dataset comprising 337 participants, 64 fashion products, and 10,816 ratings, linking affective states to fashion preferences for personalized recommendations. \\
    \midrule
    News & Ho et al. \cite{10.1145/2442810.2442816} & 3,652 articles from 21 sources, incorporating sentiment for event-based recommendations. \\
    \midrule
    Healthcare &  Serrano-Guerrero et al. \cite{SERRANOGUERRERO2024122340} & A real-world dataset of 18,952 patient reviews from ten hospitals, categorized into positive, negative, and affective reflections, annotated for sentiment analysis and hospital ranking. \\
    \midrule
    E-commerce & \vtop{\hbox{\strut Amazon Product}\hbox{\strut McAuley et al. \cite{10.1145/2766462.2767755}}\vspace{-1.2em}} & A large-scale e-commerce dataset sourced from Amazon, comprising 6 million products and over 1 million visual features.\\ 
    \midrule
    & \vtop{
    \hbox{\strut Goodreads}
    \hbox{\strut Wan et al. \cite{10.1145/3240323.3240369}}
    \vspace{-1.2em}
    } & A large-scale dataset collected from Goodreads, consisting of 229,207,408 interactions from 8,555,857 users and 2,397,688 books. \\
    \cmidrule(lr){2-3}
    Book & \vtop{\hbox{\strut Goodreads Spoiler}\hbox{\strut Wan et al. \cite{wan-etal-2019-fine}}\vspace{-1.2em}}  & A large-scale book review dataset collected from Goodreads, comprising 1,378,033 reviews across 25,475 books and 18,892 users.  \\
    \cmidrule(lr){2-3}
    & \vtop{\hbox{\strut BookCrossing}\hbox{\strut Ziegler et al. \cite{10.1145/1060745.1060754}}\vspace{-1.2em}}  & The dataset comprises 278,858 users, 1,157,112 ratings, and 271,379 books, enriched with Amazon.com’s book taxonomy of 13,525 topics and 466,573 descriptors. \\
    \midrule
    Travel & Shao et al. \cite{8796367} & The dataset consists of two sources: TripAdvisor and Trip.com. From TripAdvisor, it includes 459,180 textual comments and 43,964 images from 14,648 tourist documents, while, Trip.com contributes 293,847 textual comments and 19,492 images from 6,513 attractions.\\ 
    \midrule
    \multirow{2}{*}{Music} & \vtop{\hbox{\strut Weibo Music Emotion}\hbox{\strut Deng et al. \cite{deng2015musicemotion}}\vspace{-1.2em}} & The final dataset includes 1,059,037 records, linking 29,065 users to 59,053 music items, associating user emotions inferred from microblogs with their music preferences.\\
    \cmidrule(lr){2-3}
    & \vtop{\hbox{\strut PEIA Music}\hbox{\strut Shen et al. \cite{shen2020peia}}\vspace{-1.2em}} & The dataset includes 171,254 users, 35,993 music tracks, and 18,508,966 user-music interactions. \\
    \cmidrule(lr){2-3}
    & \vtop{\hbox{\strut TROMPA-MER}\hbox{\strut Gómez-Cañón et al. \cite{gomez2023trompa}}\vspace{-1.2em}} & The TROMPA-MER dataset includes 181 users, 691 music tracks, 4721 user-item annotations, and captures emotions using Ekman's 4 basic emotions and 7 emotions from the Geneva Emotion Music Scale. 
    \\
    \bottomrule
\end{tabular}
\end{table}
}
\normalsize

\section{Future Research Directions and Open Issues}
\label{sec:future}

Although significant progress has been made in sentiment-aware, emotion-aware, and mood-aware recommender systems, several open issues hinder the development of more robust and comprehensive affective models. The primary challenges include the lack of hybrid models that integrate multiple affective states, the scarcity of high-quality datasets for affective recommendations, the conflation of sentiment, emotion, and mood in computational models, and the need for multimodal affective learning.

\subsection{Towards Hybrid Affective Recommender Systems}

One of the most significant gaps in affective RS is the lack of hybrid models that effectively integrate multiple affective states. Existing work focuses primarily on individual affective dimensions, such as sentiment, emotion, or mood, without exploring the potential synergies between them. Sentiment analysis captures stable, long-term attitudes, emotion detection focuses on transient and intense responses, whereas mood modeling accounts for more prolonged but diffuse affective states. While these distinct types of affective states interact dynamically in human decision-making, current recommender systems have largely ignored these interactions, limiting their ability to provide a well-rounded affect-aware recommendation experience.

\begin{itemize}
    \item {\bf Unified Affective Modeling}. A promising direction for future research is to develop hybrid affective models that encode sentiment, emotion, and mood into latent spaces while capturing their interactions within a unified framework. Deep learning architectures, such as hierarchical attention networks, graph neural networks (GNNs), and transformer-based models, could be leveraged to capture the sometimes subtle relationships between different affective states. Multi-task learning could also be explored, where a single model is trained to predict multiple affective states simultaneously, improving the robustness of affective representations. Furthermore, designing adaptive recommender systems that dynamically adjust their reliance on sentiment, emotion, or mood based on the users' behavioral patterns and contextual factors could lead to more nuanced and personalized recommendations.
\end{itemize}

\subsection{Towards Broad-Coverage Affective Datasets}

The development of affective RS is severely constrained by the lack of comprehensive, large-scale datasets that include explicit affective labels. While traditional recommender systems benefit from extensive datasets containing user ratings, clicks, and purchase histories, datasets that incorporate sentiment, emotion, and mood annotations remain scarce. Many existing studies rely on manually labeled affective datasets, which are limited in scale and fail to capture the full spectrum of real-world affective states. Additionally, most affective datasets focus on only one type of affective state, making it difficult to develop hybrid models that integrate sentiment, emotion, and mood.

To address these limitations, future research includes the creation and annotation of large-scale datasets that capture a wide range of affective states. This includes data from user-generated content (e.g., reviews, tweets, blogs), physiological signals (e.g., heart rate, EEG, facial expressions), multimodal inputs (e.g., facial expressions, posture, gaze tracking, object and scene affective analysis), and contextual metadata (e.g., time of day, activity levels, and social interactions). The following techniques offer promising directions for affective dataset development:

\begin{itemize}
    \item \textbf{Weak, Distant, and Self-Supervised Learning}. These techniques help to mitigate data scarcity by generating approximate labels from unstructured data. Weak supervision can incorporate limited high-quality labeled data alongside noisy heuristic-based annotations. Distant supervision can infer affective states using external sources such as hashtags, sentiment lexicons, or emotion-tagged multimedia. Self-supervised learning can extract structured representations from unlabeled data, such as affective embeddings from video and audio interactions, which can be fine-tuned using small labeled datasets.

    \item \textbf{Transfer Learning and Domain Adaptation}. Pre-trained models can be repurposed across modalities to enhance affective modeling. For instance, facial expression recognition models trained on large video datasets can be fine-tuned for emotion-aware movie recommendations. Similarly, sentiment embeddings derived from textual reviews can be adapted to personalize image-based content, enabling seamless multimodal integration. For example, if a user’s reviews indicate a preference for uplifting, positive content, these sentiment signals could guide a recommender to suggest photos, posters, or other visual media with similar affective characteristics.

    \item \textbf{Crowdsourced Annotations}. Leveraging human annotators through platforms like Amazon Mechanical Turk or Prolific enables the scalable collection of affective labels across diverse demographics. This approach supports the creation of more representative and contextually rich datasets by incorporating subjective human judgment at scale.

    \item \textbf{Synthetic Data Generation}. Large language models and generative models (e.g., GANs) can be used to produce synthetic text, speech, or images enriched with affective content. These synthetic datasets can augment underrepresented affective categories and facilitate balanced model training.
\end{itemize}

\subsection{Addressing the Conflation of Sentiment, Emotion, and Mood}

A persistent theoretical challenge in affective RS is the conflation of sentiment, emotion, and mood. While these affective states have well-defined distinctions in psychology, many computational models fail to differentiate them adequately, leading to conceptual inconsistencies and ensuing suboptimal performance. Sentiment reflects long-term evaluative attitudes, emotions are short-lived and intense responses to specific stimuli, and moods are more diffuse affective states that persist over extended periods. However, a few existing studies blur these distinctions.
\begin{itemize}
    \item {\bf Sentiment-Emotion Conflation}. \citeauthor{wu2024bayesiansentiment} \cite{wu2024bayesiansentiment} termed a diverse set of emotions (which are short-term affective states), such as awe, anger, amusement, sadness, fear, disgust, excitement, and contentment, as sentiment (which represent a stable attitude).
    \item {\bf Emotion-Mood Conflation}. \citeauthor{yoon2012musicemotion} \cite{yoon2012musicemotion} misattributed Thayer’s mood model as a variant of the valence-arousal framework, mapping emotions such as anger, happiness, sadness, and peacefulness to musical features based on valence and arousal dimensions. In reality, Thayer’s model conceptualizes mood using orthogonal dimensions of energy and tension, which differ fundamentally from valence-arousal axes.
\end{itemize}
We recommend anchoring descriptions in a more precise vocabulary of affective terms, such as Scherer's typology of affective states\cite{scherer2005emotions} augmented with the Geneva Wheel of Emotions \cite{sacharin2012geneva}.

\subsection{Multimodal Affective Learning}

Human affective experiences are inherently multimodal, encompassing text, speech, facial expressions, physiological responses, and behavioral cues. However, most existing affective RS predominantly rely on text-based sentiment analysis or explicit user feedback, overlooking the wealth of multimodal signals that can provide a more nuanced and comprehensive understanding of user affect. This reliance on a single modality limits the accuracy and adaptability of affect-aware recommendations, as textual signal alone may not fully capture the depth and dynamics of users’ emotional and mood states. To move beyond this limitation, we highlight two key research directions:

\begin{itemize}
    \item \textbf{Building Multimodal Affective RS Frameworks}. Developing new affective recommender systems that integrate multimodal signals poses several challenges, including the fusion of heterogeneous data sources, robustness to incomplete or noisy inputs, and the need for scalable architectures capable of processing high-dimensional multimodal data. Future work should focus on constructing frameworks that combine textual, visual, auditory, and physiological inputs to improve both affective signal recognition and recommendation quality.

    \item \textbf{Using Existing Multimodal Tools and Models}. Recent advances in multimodal fusion, such as attention-based fusion networks \cite{vaswani2017attention}, cross-modal transformers \cite{yan2023crossmodal}, and contrastive learning methods \cite{khosla2020contrastive}, offer practical tools for affect modeling. Additionally, large-scale pretrained models like CLIP \cite{radford2021clip} and diffusion models \cite{larochelle2020diffusion} enable the learning of joint representations of affective states across multiple modalities. Leveraging these existing tools can enhance the robustness, adaptability, and emotional awareness of affective recommender systems beyond text-based sentiment analysis.
\end{itemize}

\section{Conclusion}

Affective recommender systems represent a significant advancement in personalization, by incorporating affective states such as sentiment, emotion, and mood. This survey presents a comprehensive and structured review of the field, grounded in a social psychology perspective of emotion, particularly Scherer's typology of affective states. Accordingly, the survey organizes affective RS publications into four main categories: attitude-aware, emotion-aware, mood-aware, and hybrid systems. Through the analysis of over 200 studies across various application domains, we have synthesized methodological trends in affective signal extraction and integration strategies, offering a unified perspective on how affect can enhance personalization in recommender systems. In doing so, we have identified 3 major research areas that are important for further progress: (1) modeling and combining different types of affective states together in affective RS; (2) the creation of large-scale datasets annotated with rich affective information, and (3) the development of robust hybrid models that account for the interplay between multiple affective dimensions.
Addressing these challenges can lead to more robust, adaptive, and human-aligned recommender systems that not only improve accuracy and engagement, but also foster emotionally intelligent and empathetic user experiences. This survey aims to provide a comprehensive resource for researchers and practitioners interested in understanding the current state of knowledge as well as in driving future innovations in affective recommender systems.

\bibliographystyle{ACM-Reference-Format}
\bibliography{arxiv-acm-csur25}


\end{document}